%% file: A20_Geometry_topology.tex
\newcommand\papertitle{\textit{Planck} 2015 results. XVIII. Background geometry and topology of the Universe}

\pdfoutput=1

\documentclass[twocolumn,traditabstract,longauth]{aa}
\usepackage{graphicx}
\usepackage{ifthen}

\usepackage[mathlines,switch]{lineno}
\modulolinenumbers[5]

\usepackage{txfonts}
\usepackage[nonamebreak]{natbib}
\usepackage{booktabs}
\usepackage[usenames,dvipsnames,svgnames,table]{xcolor}

\usepackage[labelfont=bf]{caption}
\usepackage{subcaption}   
\usepackage{fixltx2e}

\usepackage{hyperref}
\hypersetup{breaklinks,colorlinks,citecolor=blue}

\newcommand{\rev}[1]{{{#1}}}

\bibpunct{(}{)}{;}{a}{}{,}  
\input Planck.tex
\input macros_data.tex

\newcommand{\eqref}[1]{(\ref{#1})}

\newcommand{\tbl}[1]{Table~#1}

\newcommand{\fig}[1]{Fig.~#1}

\newcommand{\ie}{\mbox{i.e.}}

\newcommand{\bianchiviih}{{Bianchi VII{$_{\lowercase{h}}$}}}

\newcommand{\vect}[1]{\ensuremath{\mbox{\boldmath ${#1}$}}}

\newcommand{\Den}{\ensuremath{\Omega}}

\newcommand{\euls}{\ensuremath{\eula, \eulb, \eulc}}
\newcommand{\eula}{\ensuremath{\alpha}}
\newcommand{\eulb}{\ensuremath{\beta}}
\newcommand{\eulc}{\ensuremath{\gamma}}

\newcommand{\bx}{\ensuremath{x}}

\newcommand{\bparam}{\ensuremath{\Theta_{\rm B}}}
\newcommand{\cosmoparam}{\ensuremath{\Theta_{\rm C}}}
\newcommand{\fitdata}{\ensuremath{\vec{d}}}

\newcommand{\elmax}{\ensuremath{{\ell_{\rm max}}}}

\newcommand{\mtrx}[1]{\ensuremath{\tens{{#1}}}}
\newcommand{\nilc}{{\tt NILC}}
\newcommand{\smica}{{\tt SMICA}}
\newcommand{\sevem}{{\tt SEVEM}}
\newcommand{\commander}{{\tt Commander}}
\newcommand{\anicosmo}{{\tt AniCosmo}}
\newcommand{\multinest}{{\tt MultiNest}}

\begin{document}
\input{A20_Geometry_topology_authors_and_institutes.tex}


\title{\papertitle}

\abstract{

  Maps of cosmic microwave background (CMB) temperature and polarization from the 2015 release of \Planck\ data provide the
  highest-quality full-sky view of the surface of last scattering available to date. This allows us
  to detect possible departures from a globally isotropic
  cosmology.
  We present the first searches using CMB polarization for correlations induced by a possible non-trivial
  topology with a fundamental domain intersecting, or nearly intersecting, the last-scattering
  surface (at comoving distance ${\chi_\mathrm{rec}}$), both via a direct scan for matched circular
  patterns at the intersections and by an optimal likelihood calculation for specific topologies.
  We specialize to flat spaces with cubic toroidal (T3) and slab (T1) topologies, finding that explicit searches for the latter
  are sensitive to other topologies with antipodal symmetry.
  These searches yield no detection of a
  compact topology with a scale below the diameter of the last-scattering surface.
  The limits on the radius
  ${\cal R}_{\rm i}$ of the largest sphere inscribed in the fundamental domain (at log-likelihood ratio
  $\Delta\ln{\cal L}> -5$ relative to a simply-connected flat \Planck\ best-fit model) are:
  ${\cal R}_\mathrm{i}>0.97\,\chi_\mathrm{rec}$ for the T3 cubic torus; and
  ${\cal R}_\mathrm{i}>0.56\,\chi_\mathrm{rec}$ for the T1 slab.
  The limit for the T3 cubic torus from the matched-circles search is numerically equivalent,
  ${\cal R}_\mathrm{i}>0.97\,\chi_\mathrm{rec}$ at 99\,\% confidence level from polarization data alone.
  We also perform a Bayesian search for an anisotropic global
  \bianchiviih\ geometry. In the
  non-physical setting where the Bianchi cosmology is decoupled from the standard cosmology, \Planck\ temperature
  data favour the inclusion of a Bianchi component with a Bayes factor of at least 2.3 units of
  log-evidence. However, the cosmological parameters that generate
  this pattern are in strong disagreement with those found from CMB anisotropy data alone.
  Fitting the induced polarization pattern for this model to the \Planck\ data requires an amplitude of $-0.10\pm0.04$ compared to the
  value of $+1$ if the model were to be correct.
  In the physically motivated setting where the Bianchi parameters are coupled and fitted simultaneously
  with the standard cosmological parameters, we find no evidence for a \bianchiviih\ cosmology and
  constrain the vorticity of such models to $(\omega/H)_0 < 7.6 \times 10^{-10}$ (95\,\% CL).}


\titlerunning{\papertitle}
\authorrunning{Planck Collaboration}

\maketitle
\alltwentyfifteenresultspapers

\section{Introduction} 
\label{sec:introduction}


This paper, one of a series associated with the 2015 release of \Planck\footnote{\Planck\ (\url{http://www.esa.int/Planck}) is a project of the European Space Agency  (ESA) with instruments provided by two scientific consortia funded by ESA member states and led by Principal Investigators from France and Italy, telescope reflectors provided through a collaboration between ESA and a scientific consortium led and funded by Denmark, and additional contributions from NASA (USA).}\ data, will present limits on departures from the global isotropy of spacetime.
We will assess anisotropic but homogeneous Bianchi cosmological models and non-trivial global topologies in the light of the latest temperature and polarization data.

In \citet{planck2013-p19}, the limits came from the 2013 \Planck\ cosmological data release: cosmic microwave background (CMB) intensity data collected over approximately one year. This work will use the 2015 \Planck\ data: CMB intensity from the whole mission along with a subset of polarization data.
The greater volume of intensity data will allow more restrictive limits on the possibility of topological scales slightly larger than the volume enclosed by the last-scattering surface (roughly the Hubble volume), probing the excess anisotropic correlations that would be induced at large angular scales were such a model to obtain. For cubic torus topologies, we can therefore observe explicit repeated patterns (matched circles) when the comoving length of an edge is less than twice the distance to the recombination surface, $\chi_\mathrm{rec}\simeq3.1H_0^{-1}$ (using units with $c=1$ here and throughout).
Polarization, on the other hand, largely generated during recombination itself, can provide a more sensitive probe of topological domains smaller than the Hubble volume.

Whereas the analysis of temperature data in multiply connected
universes has been treated in some depth in the literature
\citep[see][and references therein]{planck2013-p19}, the discussion of
polarization has been less complete.
This paper will therefore extend our previous likelihood analysis to polarized data, update the direct search for matched circles \citep{cornish2004} as discussed in \citet{bielewicz2012}, and use these to present the first limits on global topology from polarized CMB data.

The cosmological properties of Bianchi models \citep{collins:1973, barrow:1985}, were initially discussed in the context of CMB intensity \citep{barrow:1986, jaffe:2006b, jaffe:2006c, pontzen:2009}.  As discussed in \citet{planck2013-p19} it is by now well known that the observed large-scale intensity pattern mimics that of a particular Bianchi VII$_h$ model, albeit one with cosmological parameters quite different from those needed to reproduce other CMB and cosmological data.  More recently the induced polarization patterns have been calculated \citep{pontzen:2007, pontzen:2009, pontzen:2011}. In this paper, we analyse the complete \Planck\ intensity data, and compare the polarization pattern induced by that anisotropic model to \Planck\ polarization data.

We note that the lack of a strong detection of cosmic $B$-mode polarization already provides some information about the Bianchi models: the induced geometrical focusing does not distinguish between $E$ and $B$ and thus should produce comparable amounts of each \citep[e.g.,][]{pontzen:2009}.
Note that this does \emph{not} apply to topological models: the linear evolution of primordial perturbations guarantees that a lack of primordial tensor perturbations results in a lack of $B$-mode polarization---the transfer function is not altered by topology.

In Sect.~\ref{sub:previous_results}, we discuss previous limits on anisotropic models from \Planck\ and other experiments. In Sect.~\ref{sec:cmb_polarization_signals_in_anisotropic_and_multiply_connected_universes} we discuss the CMB signals generated in such models, generalized to both temperature and polarization. In Sect.~\ref{sec:methods} we describe the \Planck\ data and simulations we use in this study, the different methods we apply to those data, and the validation checks performed on those simulations. In Sect.~\ref{sec:results} we discuss the results and conclude in Sect.~\ref{sec:discussion} with the outlook for application of these techniques to future data and broader classes of models.



\section{Previous results} 
\label{sub:previous_results}

The first searches for non-trivial topology on cosmic scales looked for
repeated patterns or individual objects in the distribution of
galaxies
\citep{Shvartsman1974,FangSato1983,Fagundes:1987dn,1996A&A...313..339L,1996MNRAS.283.1147R,Weatherley:2003gw,2011A&A...529A.121F}. Searches for topology using the CMB began with COBE \citep{bennett:1996} and found no indications of a non-trivial topology on the scale of the last-scattering surface (e.g., \citealt{Starobinskii1993,Sokolov1993,StevensScottSilk1993,deoliveira-costa1995,Levin:1998bw,
BPS1998,BPS2000b,Rocha2004}; but see also \citealt{2000MNRAS.312..712R,2000CQGra..17.3951R}). With the higher resolution and sensitivity of WMAP, there were indications of low power on large scales which could have had a topological origin \citep{jarosik2010,Luminet:2003bp,Caillerie:2007jv,1999ApJ...524..497A,aurich2004,aurich2005,aurich2006,aurich2008,aurich2013,lew2008,roukema2008,Niarchou:2003gi}, but this possibility was not borne out by detailed real- and harmonic-space analyses in two dimensions \citep{cornish2004,key2007,bielewicz2009,Dineen2005,Kunz:2005wh,phillips2006,Niarchou:2007fe}. Most studies, including this work, have emphasized searches for fundamental domains with antipodal correlations; see \rev{\citet{Vaudrevange2012} for results from a general search for the patterns induced by non-trivial topology on scales within the volume defined by the last-scattering surface, and,} e.g., \citet{aurich2014} for a recent discussion of other possible topologies.

For a more complete overview of the field, we direct the reader to \citet{planck2013-p19}.
In that work, we applied various techniques to the \Planck\ 2013 intensity data. For topology, we showed that a fundamental topological domain smaller than the Hubble volume is strongly disfavoured. This was done in two ways: first, a direct likelihood calculation of specific topological models; and second, a search for the expected repeated ``circles in the sky'' \citep{cornish2004}, calibrated by simply-connected simulations. Both of these showed that the scale of any possible topology must exceed roughly the distance to the last-scattering surface, $\chi_\mathrm{rec}$. For the cubic torus, we found that the radius of the largest sphere inscribed in the topological fundamental domain must be ${\cal R}_\mathrm{i}>0.92\,\chi_\mathrm{rec}$  (at log-likelihood ratio
  $\Delta\ln{\cal L}> -5$ relative to a simply-connected flat \Planck\ 2013 best-fit model). The matched-circle limit on topologies predicting back-to-back circles was ${\cal R}_\mathrm{i}>0.94\,\chi_\mathrm{rec}$ at the
    99\,\% confidence level.

Prior to the present work, there have been some extensions of the search for cosmic topology to polarization data.
In particular, \citet{bielewicz2012} \rev{\citep[see also][]{riazuelo2006}} extended the direct search for matched circles to polarized data and found that the available WMAP data had insufficient sensitivity to provide useful constraints.

For \bianchiviih\ models, in \citet{planck2013-p19} a full Bayesian analysis
of the \Planck\ 2013 temperature data was performed, following the methods of \citet{mcewen:bianchi}.  It was
concluded that a physically-motivated model was not
favoured by the data. If considered as a phenomenological template (for which the parameters common to the standard stochastic CMB and the deterministic \bianchiviih\ component are not linked),
it was shown that an unphysical \bianchiviih\ model is favoured, with
a log-Bayes factor between $1.5\pm0.1$ and $2.8\pm0.1$---equivalent to
an odds ratio of between approximately 1:4 and 1:16---depending of
the component separation technique adopted.
Prior to the analysis of \citet{planck2013-p19}, numerous analyses of
Bianchi models using COBE \citep{bennett:1996} and {WMAP}
\citep{jarosik2010} data had been performed \citep{bunn:1996,
  kogut:1997, jaffe:2005, jaffe:2006c, jaffe:2006b, jaffe:2006a,
  cayon:2006, lm:2006, mcewen:2006:bianchi, bridges:2006b, ghosh:2007,
  pontzen:2007, bridges:2008, mcewen:bianchi}, and a similar Bianchi
template was found in the {WMAP} data, first by \citet{jaffe:2005} and
then subsequently by others \citep{bridges:2006b, bridges:2008,
  mcewen:bianchi}.  \citet{pontzen:2007} discussed the CMB
polarization signal from Bianchi models, and showed some
incompatibility with {WMAP} data due to the large amplitude of both $E$-
and $B$-mode components.  For a more detailed review of the analysis of
Bianchi models we refer the reader to \citet{planck2013-p19}.


\section{CMB signals in anisotropic and multiply-connected universes} 
\label{sec:cmb_polarization_signals_in_anisotropic_and_multiply_connected_universes}

\subsection{Topology} 
\label{sub:topology}

There is a long history of studying the possible topological compactification of Friedmann-Lema\^itre-Robertson-Walker (FLRW) cosmologies; we refer readers to overviews such as \citet{Levin:2002co}, \citet{LachiezeRey:1995wi}, and \citet{riazuelo2004a,riazuelo2004b} for mathematical and physical detail. The effect of a non-trivial topology is equivalent to considering the full ({simply-connected}) three dimensional spatial slice of the manifold (the {covering space}) as being tiled by identical repetitions of a shape which is finite in one or more directions, the {fundamental domain}. In flat universes, to which we specialize here, there are a finite number of possibilities, each described by one or more continuous parameters describing the size in different directions.

In this paper, we pay special attention to topological models in which the fundamental domain is a right-rectangular prism (the three-torus, also referred to as ``T3''), possibly with one or two infinite dimensions (the T2 ``chimney'' or ``rod'', and T1 ``slab'' models). We limit these models in a number of ways. We explicitly compute the likelihood of the length of the fundamental domain for the cubic torus. Furthermore, we consider the slab model as a proxy for other models in which the matched circles (or excess correlations) are antipodally aligned, similar to the ``lens'' spaces available in manifolds with constant positive curvature. These models are thus sensitive to tori with varying side lengths, including those with non-right-angle corners. In these cases, the likelihood would have multiple peaks, one for each of the aligned pairs; their sizes correspond to those of the fundamental domains and their relative orientation to the angles. These non-rectangular prisms will be discussed in more detail in \citet{JaffeStarkman2015}.

\subsubsection{Computing the covariance matrices}
\label{subsub:corr_mat_calc}


In \citet{planck2013-p19} we computed the temperature-temperature ($TT$) covariance matrices by
summing up all modes $\vec{k}_{\vec{n}}$ that are present given the boundary conditions imposed
by the non-trivial topology. For a cubic torus, we have a three-dimensional wave vector $\vec{k}_{\vec{n}} = (2 \pi / L) \vec{n}$ for a triplet of integers $\vec{n}$, with unit vector ${\vec{\hat{k\,}}}$ and the harmonic-space covariance matrix
\begin{equation}\label{topocorrT}
C_{\ell \ell'}^{m m' \, (TT)}
\propto \sum_{\vec{n}}  \Delta_{\ell}^{(T)}(k_n, \Delta\eta) \Delta_{\ell'}^{(T)}(k_n, \Delta\eta) P(k_n) Y_{\ell m}({\vec{\hat{k\,}}})
Y_{\ell' m'}^* ({\vec{\hat{k\,}}}) \, ,
\end{equation}
where $\Delta_\ell^{(T)} (k, \Delta\eta)$ is the temperature radiation transfer
function (see, e.g., \citealt{Bond:1987uc} and \citealt{Seljak1996cmbfast}).

It is straightforward to extend this method to include polarization, since the cubic topology affects neither the local
physics that governs the transfer functions, nor the photon propagation. The only effect is the discretization of the modes.
We can therefore simply replace the radiation transfer function for the temperature fluctuations with the one for
polarization, and obtain
\begin{equation}\label{topocorrX}
C_{\ell \ell'}^{m m' \, (XX')}
\propto \sum_{\bf n}  \Delta_{\ell}^{(X)}(k_n, \Delta\eta) \Delta_{\ell'}^{(X')}(k_n, \Delta\eta) P(k_n) Y_{\ell m}({\vec{\hat{k\,}}})
Y_{\ell' m'}^* ({\vec{\hat{k\,}}}) \, ,
\end{equation}
where $X,X' = E,T$. We are justified in ignoring the possibility of $B$-mode polarization as it is sourced only by primordial gravitational radiation even in the presence of non-trivial topology.
In this way we obtain three sets of covariance matrices: $TT$, $TE$, and $EE$. In addition, since the publication of \citet{planck2013-p19} we have optimized
the cubic torus calculation by taking into account more of the symmetries. The resulting speed-up of about an order of magnitude
allowed us to reach a higher resolution of $\ell_\mathrm{max} = 64$.

The fiducial cosmology assumed in the calculation of the covariance matrices is a flat $\Lambda$CDM FLRW universe with
Hubble constant $H_0=100h\,\textrm{km}\,\textrm{s}^{-1}\,\textrm{Mpc}^{-1}$, where: $h=0.6719$; scalar spectral index $n_\mathrm{s}=0.9635$;
baryon density $\Omega_\mathrm{b}h^2=0.0221$;
cold dark matter density $\Omega_\mathrm{c}h^2=0.1197$; and
neutrino density $\Omega_\nu h^2=0.0006$.

\subsubsection{Relative information in the matrices}
\label{sec:KL}

To assess the information content of the covariance matrices, we consider the
Kullback-Leibler (KL) divergence (see, e.g., \citealt{Kunz:2005wh,Kunz:2008fq}
and \citealt{planck2013-p19} for further applications of the KL divergence to topology).
The KL divergence
between two probability distributions $p_1(x)$ and $p_2(x)$ is given
by
\begin{equation}
    d_\mathrm{KL} = \int p_1(x)\ln\frac{p_1(x)}{p_2(x)}\;dx\;.
\end{equation}
If the two distributions are Gaussian with covariance matrices $\mtrx{C}_1$
and $\mtrx{C}_2$, this expression simplifies to
\begin{equation}
    d_\mathrm{KL} = -\frac12 \left[ \ln\left|\mtrx{C}_1 \mtrx{C}_2^{-1}\right| + \mathrm{Tr}\left(\mtrx{I} - \mtrx{C}_1 \mtrx{C}_2^{-1}\right)\right]\;,
\end{equation}
and is thus an asymmetric measure of the discrepancy between the covariance
matrices. The KL divergence can be interpreted as the ensemble
average of the log-likelihood ratio $\Delta\ln{\cal L}$ between realizations of the two distributions.
Hence, it enables us to probe the ability to tell if, on average, we can distinguish
realizations of $p_1$ from a fixed $p_2$ without having to perform a brute-force Monte Carlo integration.
Thus, the KL divergence is related to ensemble averages of the likelihood-ratio plots that we present for simulations (Sect.~\ref{par:likelihood_validation}) and real data (Sect.~\ref{sub:results_topology}) but can be calculated from the covariance matrices alone. Note that with this definition, the KL divergence is \emph{minimized} for cases with the best match (\emph{maximal} likelihood).

In \citet{planck2013-p19} we used the KL divergence to show that the likelihood is robust to differences in the cosmological model and small differences in the topology.

\begin{figure}[htbp]
  \centering
  \includegraphics[width=1.0\columnwidth]{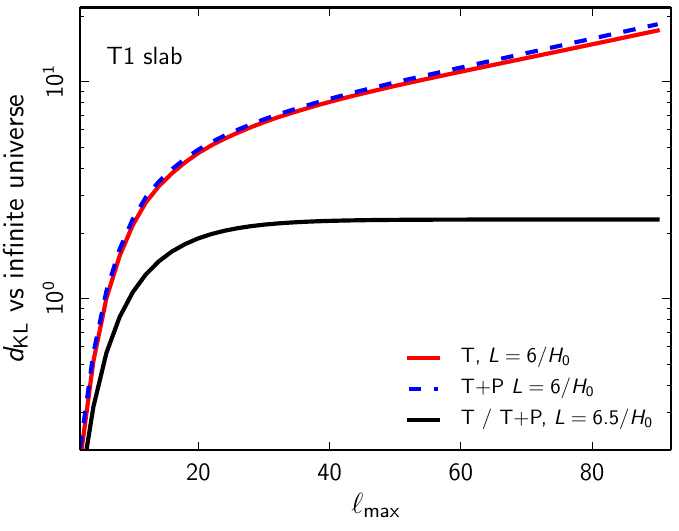}
  \includegraphics[width=1.0\columnwidth]{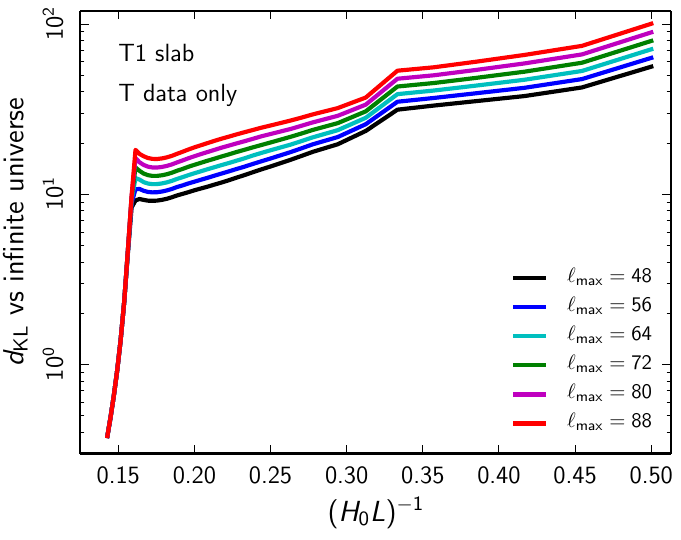}
  \caption{The Kullback-Leibler divergence of slab (T1) topologies relative to an infinite universe as a function of $\ell_\mathrm{max}$ with sizes $L=6H_0^{-1}$ and $L=6.5H_0^{-1}$ (top), and as a function of size $L$ of the fundamental domain for various $\ell_\mathrm{max}$  (bottom). \rev{A torus with $L>6.2 H_0^{-1}$, corresponding to $(H_0 L)^{-1} < 0.154$, has a fundamental domain that is larger than the distance to the last-scattering surface and leaves only a small trace in the CMB. This is why the KL divergence drops rapidly at this point.}
Note that the information for $L=6H_0^{-1}$
 continues to rise with $\ell_\mathrm{max}$ whereas it levels off for the slightly larger $L=6.5H_0^{-1}$ case. \rev{In the lower
 panel we see that there is a slight feature in $d_\mathrm{KL}$ at about half the horizon distance, which is probably due to harmonic effects.}
 The corresponding figures for cubic (T3) topologies look
qualitatively similar except that all $d_\mathrm{KL}$ values are three times larger.}
\label{fig:topo_kl_lmax}
\end{figure}

In Fig.~\ref{fig:topo_kl_lmax} we plot the KL divergence relative to an infinite universe for the slab topology as a function of resolution $\ell_\mathrm{max}$ (upper panel) and fundamental domain size (lower panel).
Our ability to detect
a topology with a fundamental domain smaller than the distance to the last-scattering surface (approximately at the horizon distance $\chi_\mathrm{rec}=3.1H_0^{-1}$, so with sides of length $L=2\,\chi_\mathrm{rec}=6.2H_0^{-1}$)
grows significantly with the resolution even beyond the cases
that we studied. For the noise levels of the 2015 \lowEB\ data considered here and defined in \citet{planck2014-a15},
polarization maps do not add much information beyond that contained in the temperature maps, although, as also shown in Sect.~\ref{par:likelihood_validation}, the higher sensitivity achievable by the full \Planck\ low-$\ell$ data over all frequencies should enable even stronger constraints on these small fundamental domains.

If, however, the fundamental domain is larger than the horizon (as is the case for $L=6.5H_0^{-1}$) then the relative information in the
covariance matrix saturates quite early and a resolution of $\ell_\mathrm{max}\simeq48$ is actually sufficient. The main goal is thus to ensure that we have enough
discriminatory power right up to the horizon size.
In addition, polarization
does not add much information in this case, irrespective of the noise level. This is to be expected: polarization is generated only for a short period of time around the surface of last scattering. Once the fundamental domain exceeds the horizon size, the relative information drops rapidly towards
zero, and the dependence on $\ell_\mathrm{max}$ becomes weak.

\begin{figure}[htbp]
  \centering
  \includegraphics[width=1.0\columnwidth]{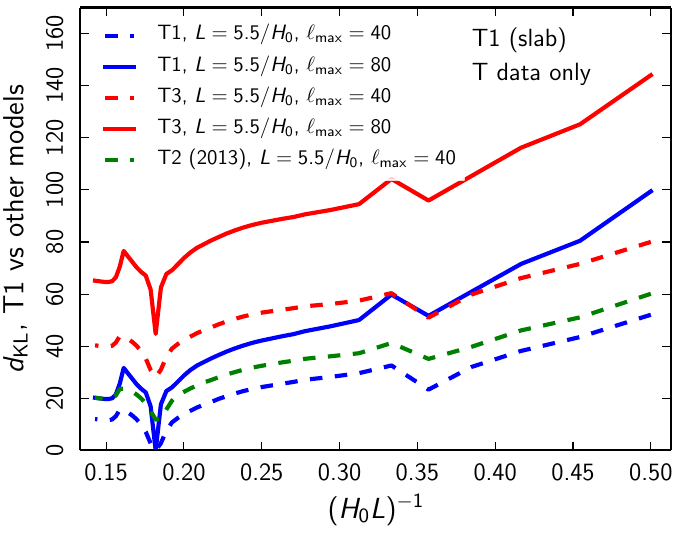}
  \caption{The Kullback-Leibler divergence of fixed cubic, rod, and slab topologies with fundamental domain side $L=5.5H_0^{-1}$ compared to a slab of variable fundamental domain size $L$. The chimney space T2 dates from the 2013 analysis
  \citep{planck2013-p19} and was computed for the best-fit parameters of that release. \rev{In all cases the smallest KL
  divergence, corresponding to the best fit, appears at $L=5.5 H_0^{-1}$, indicating that the slab space can be used to detect other topologies.
  An additional dip at $L\simeq 5.5/(2 H_0)$ may be due to a harmonic effect at half the size of the fundamental domain; it is, however, much smaller than the drop in KL divergence at the size of the fundamental domain.}}
\label{fig:topo_kl_X}
\end{figure}

In Fig.~\ref{fig:topo_kl_X} we plot the KL divergence as a function of the size of the fundamental domain for fixed cube (T3), rod (T2), and slab (T1) topologies, each with fundamental domain size $L=5.5H_0^{-1}$, compared to the slab. Each shows a strong dip at $L=5.5H_0^{-1}$, indicating the ability to detect this topology (although note the presence of a weaker dip around half the correct size, $L\simeq2.75H_0^{-1}$). The figure also shows that $\ell_\mathrm{max}=40$ still shows the dip at the correct location, although somewhat more weakly than $\ell_\mathrm{max}=80$.

\begin{figure}[htbp]
  \centering
  \includegraphics[width=1.0\columnwidth]{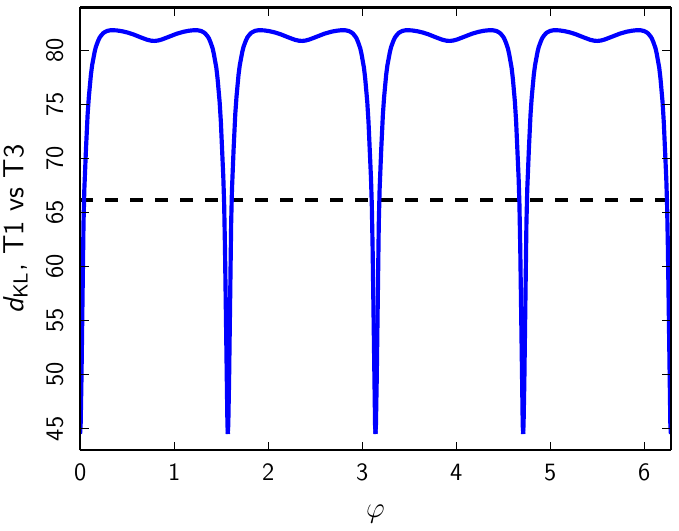}
  \caption{The Kullback-Leibler divergence of a slab space relative to a cubic topology, as a function of rotation angle of the slab space (blue curve). Both spaces have $L=5.5H_0^{-1}$ and $\ell_\mathrm{max}=80$. The horizontal black dashed line gives the Kullback-Leibler divergence of an infinite universe relative to the cubic topology and illustrates how much better the slab space fits with the correct orientation relative to the cubic torus.}
\label{fig:topo_kl_rot}
\end{figure}

Note that the shape of the curves is essentially identical, with the slab likelihood able to detect one or more sets of antipodal matched circles (and their related excess correlations at large angular scales) present in each case. Figure~\ref{fig:topo_kl_X} therefore shows that using the covariance matrix for a slab (T1) topology also allows detection of rod (T2) and cubic (T3) topologies: this is advantageous as the slab covariance matrix is considerably easier to calculate than the cube and rod, since it is only discretized in a single direction. Figure~\ref{fig:topo_kl_rot} shows the KL divergence as a function of the relative rotation of the fundamental domain, showing that, despite the lack of the full set of three pairs of antipodal correlations, we can determine the relative rotation of a single pair.  This is exactly how the matched-circles tests work. Furthermore, as we will demonstrate in Sect.~\ref{par:likelihood_validation}, slab likelihoods are indeed separately sensitive to the different sets of antipodal circles in cubic spaces. We can hence adopt the slab as the most general tool for searching for spaces with antipodal circles.

\subsection{Bianchi models} 
\label{sub:bianchi_models}


The polarization properties of Bianchi models were first derived
in \citet{pontzen:2007} and extensively categorized in
\citet{pontzen:2009} and \citet{pontzen:2011}. In these works it was shown that advection in Bianchi
universes leads to efficient conversion of $E$-mode polarization to
$B$ modes; evidence for a significant Bianchi component found in
temperature data would therefore suggest a large $B$-mode signal (but
not necessarily require it; see \citealt{pontzen:2009}). For examples of the
temperature and polarization signatures of \bianchiviih\ models we
refer the reader to figure 1 of \citet{pontzen:2009}.
Despite the potential for CMB polarization to constrain the Bianchi sector, a full
polarization analysis has not yet been carried out. The analysis
of \citet{pontzen:2007} remains the state-of-the-art, where WMAP $BB$ and
$EB$ power spectra were used to demonstrate (using a simple $\chi^2$
analysis) that a \bianchiviih\ model derived from temperature data was
disfavoured compared to an isotropic model.

The subdominant, deterministic CMB contributions of \bianchiviih\
models can be characterized by seven parameters: the matter and
dark energy densities, $\Den_{\rm m}$ and $\Den_{\Lambda}$,
respectively; the present dimensionless vorticity, $(\omega/H)_0$; the
dimensionless length-scale parameter, $x$, which controls the
``tightness'' of the characteristic Bianchi spirals; and the Euler
angles\footnote{The active $zyz$
  Euler convention is adopted, corresponding to the rotation of a
  physical body in a {\em{fixed}\/} coordinate system about the $z$, $y$,
  and $z$ axes by $\gamma$, $\beta$, and $\alpha$, respectively.},
$(\euls)$, describing their orientation (\ie, the choice of coordinate
system), where $H$ is the Hubble parameter. For further details see
\citet{planck2013-p19}, \citet{mcewen:bianchi}, \citet{pontzen:2009},
\citet{pontzen:2007}, \citet{jaffe:2006b}, \citet{jaffe:2005}, and \citet{barrow:1985}.



\section{Methods} 
\label{sec:methods}

\subsection{Data} 
\label{sub:data}

In this work we use data from the \Planck\ 2015 release. This includes intensity maps from the full mission, along with a subset of polarization data. Specifically, for the likelihood calculations discussed below (Sect.~\ref{subs:topology} for application to topology and Sect.~\ref{subs:bianchi} for Bianchi models) which rely on {\tt HEALPix} maps at $N_{\rm
  side}=16$, we use the data designated ``\lowTP'', as defined for the low-$\ell$ \Planck\ likelihood for isotropic models \citep{planck2014-a13,planck2014-a15}: \lowEB\ polarization maps  based on the LFI 70\GHz\ channel and lowT temperature maps created by the \commander\ component separation method, along with the appropriate mask and noise covariance matrix. As in \citet{planck2013-p19}, the intensity noise contribution is negligible on these scales, and diagonal regularizing noise with variance $\sigma_I^2 = 4\,\mu{\rm K}^2$ has therefore been added to the intensity portion of the noise covariance matrix.
We cut contaminated regions of the sky using the low-$\ell$ mask defined for the \Planck\ isotropic likelihood code \citep{planck2014-a13}, retaining 94\,\% of the sky for temperature, and the \lowTP\ polarization mask, cleaned with the templates created from \Planck\ 30\GHz\ and 353\GHz\ data, retaining 47\,\% of the sky for polarization.

The matched-circle search (Sects.~\ref{sec:method_circles}
and~\ref{ssub:matched_circles}) uses four component-separated maps
\citep{planck2014-a11} which effectively combine both intensity and
polarization information from different scales. The maps are smoothed
with a Gaussian filter of 30\arcm\ and 50\arcm\ full width at
half maximum (FWHM) for temperature and polarization, respectively, and
degraded to $N_\mathrm{side}=512$. Corresponding temperature and
polarization common masks for diffuse emission, with a point
source cut for the brightest sources, downgraded analogously to the maps, are
used. After degradation, and accounting for the needed expansion of
the polarization mask due to the conversion of $Q$ and $U$ to $E$, the
temperature map retains 74\,\% of the sky and the polarization map
40\,\%. These $E$-mode maps are calculated using the method of
\citet{bielewicz2012} \citep[see also][]{kim2011} and correspond to
the spherical Laplacian of the scalar $E$, consequently filtering out
power at large angular scales.



\subsection{Topology: matched circles} 
\label{sec:method_circles}


As in \citet{planck2013-p19}, we use the
circle comparison statistic of \citet{cornish1998}, optimized for small-scale
anisotropies \citep{cornish2004}, to search for correlated circles
in sky maps of the CMB temperature and polarization anisotropy.
\rev{The circle comparison statistic uses the fact that the intersection of the
  topological fundamental domain with the surface of last scattering
  is a circle, potentially viewed from different
  directions in a multiply-connected universe. Contrary to the temperature
  anisotropy, sourced by multiple terms at the last-scattering surface (i.e., the internal photon
  density fluctuations combined with the ordinary Sachs-Wolfe and Doppler effects), the CMB polarization anisotropy is sourced only by the quadrupole
  distribution of radiation scattering from free electrons at the moment
  of recombination \citep[e.g.,][]{Kosowsky:1996kc}. In particular, the recombination signal from polarization is only generated for a short time while there are enough electrons to scatter the photons but few enough for the plasma to be sufficiently transparent.
  Thus, in a multi-connected universe the polarization signal
  does not exhibit the same cancellation of contributions
 from different terms as in the temperature
 anisotropy \citep{bielewicz2012}.
 Polarization thus can provide a
better opportunity for the detection of topological
signatures than a temperature anisotropy map.}
There is a small subtlety here: whereas the intensity is a scalar
and thus is unchanged when viewed from different directions, the
polarization is a tensor which behaves differently under
rotation. The polarization pattern itself depends on the viewing angle; hence, we need to use the coordinate-independent quantities,
$E$ and $B$, which are scalars (or pseudo-scalars) and are thus
unchanged when viewed from different directions.

The decomposition into $E$ and $B$ of an arbitrary masked CMB
polarization map, contaminated by noise, foregrounds, and systematic
errors, is itself a computationally demanding task, non-local on the sky.
Assuming negligible initial $B$ polarization, we use only the $E$ maps
produced \rev{from component-separated CMB polarization maps using
the same approach as \citet{bielewicz2012}}.

\rev{Compared with the likelihood method described below, the circles
search uses higher-resolution maps, and thus is sensitive out to
a much higher maximum multipole, $\ell_\mathrm{max}$. It is also potentially less
sensitive to large-scale systematic errors, as the lowest
multipoles are effectively filtered out:
the polarization signal is weighted by a factor proportional to
$\ell^2$ in the transformation from the Stokes parameters $Q$ and $U$
to an $E$-mode map. From the results of Sect.~\ref{sec:KL}, this
indicates that it uses more of the information available when
confronting models with fundamental domains within the
last-scattering surface compared to our implementation of the
likelihood, limited to $\ell_\mathrm{max}\simeq40$. As we show in
Sect.~\ref{sec:applic_simulations}, this also allows the use of
high-pass filtered component-separated maps \citep[as defined
in][]{planck2014-a11} without a significant decrease in the ability
to detect a multiply-connected topology.}

The matched-circle statistic is defined by
\begin{equation}\label{eqn:s_statistic_fft}
S_{i,j}^{+}(\alpha, \phi_\ast)=\frac{2 \sum_m |m|\, X_{i,m}^{}
   X_{j,m}^\ast
e^{-{\rm i} m \phi_\ast}}{\sum_n |n| \left( |X_{i,n}|^2+|X_{j,n}|^2\right)}\ ,
\end{equation}
where $X_{i,m}$ and $X_{j,m}$ denote the Fourier
coefficients of the temperature or $E$-mode fluctuations around two circles of angular
radius $\alpha$ centred at different points on the sky, $i$ and $j$,
respectively, with relative phase $\phi_\ast$. The $m$th harmonic of
the field anisotropies around the circle is weighted by the factor $|m|$,
taking into account the number of degrees of freedom per mode. Such
weighting enhances the contribution of small-scale structure relative to large-scale
fluctuations.

The $S^{+}$ statistic corresponds to pairs of circles, with the points
ordered in a clockwise direction (phased). For
alternative ordering, when the points are
ordered in an anti-clockwise direction (anti-phased along one of the circles), the
Fourier coefficients $X_{i,m}$ are complex conjugated, defining the $S^{-}$ statistic. This
allows the detection of both orientable and non-orientable
topologies. For orientable topologies the matched circles have
anti-phased correlations, while for non-orientable topologies they have
a mixture of anti-phased and phased correlations.

The $S^\pm$ statistics take values over the interval
$[-1,1]$. Circles that are perfectly matched have $S=1$, while
uncorrelated circles will have a mean value of $S=0$.
To find matched circles for each radius $\alpha$, the maximum value
\mbox{$S_{\rm max}^{\pm}(\alpha) ={\rm max}_{i,j,\phi_\ast} \, S_{i,j}^{\pm}(\alpha,\phi_\ast)$}
is determined.

Because general searches for matched circles are computationally very
intensive, we restrict our analysis to a search for pairs of
circles centred around antipodal points, so called back-to-back
circles. The maps are also downgraded as described in Sect.~\ref{sub:data}.
This increases the signal-to-noise ratio and greatly speeds up the
computations required, but with no significant loss of discriminatory
power. Regions most contaminated by Galactic foreground were removed from the
analysis using the common temperature or polarization mask. More details
on the numerical implementation of the algorithm can be found in
\citet{bielewicz2011} and \citet{bielewicz2012}.

To draw any conclusions from an analysis based on the statistic $S_{\rm
  max}^{\pm}(\alpha)$, it is very important to correctly estimate the
threshold for a statistically significant match of circle pairs.
We used 300 Monte Carlo simulations of \rev{the \Planck\ \smica\ maps
  processed in the same way as the data}
to establish the threshold such that fewer than
1\,\% of simulations would yield a false event. Note that we perform
the entire analysis, including the final statistical calibration,
separately for temperature and polarization.



\subsection{Likelihood} 
\label{sub:likelihood}


\subsubsection{Topology} 
\label{subs:topology}

For the likelihood analysis of the large angle intensity and polarization data we
have generalized the method implemented in \citet{planck2013-p19} to include polarization.
The likelihood, i.e., the probability to find a combined temperature and polarization data map
$\vec{ d}$ with associated noise matrix $\mtrx{ N}$ given a certain
topological model $T$ is then given by
\begin{eqnarray}\label{eq:fullskylike}
\lefteqn{P(\vec{d}| \mtrx{C}[ \Theta_\mathrm{C},\Theta_\mathrm{T},T],A,\varphi)}\nonumber \\
&& \propto \frac{1}{\sqrt{|A \mtrx{C} + \mtrx{ N}|}}
\exp\left[ - \frac{1}{2} \vec{d}^* (A \mtrx{ C} + \mtrx{ N} )^{-1} \vec{d} \right] \, ,
\end{eqnarray}
where now $\vec{d}$ is a $3 N_\mathrm{pix}$-component data vector obtained by concatenation
of the $(I,Q,U)$ data sets while $\mtrx{C}$ and $\mtrx{N}$ are
$3 N_\mathrm{pix} \times 3 N_\mathrm{pix}$ theoretical signal and noise covariance matrices,
arranged in the block form as
\begin{equation}\label{eq:polcorrmat}
\mtrx{C} =  \left(
\begin{array}{ccc}
C_{II} & C_{IQ} & C_{IU}\\
C_{QI} & C_{QQ} & C_{QU}\\
C_{UI} & C_{UQ} & C_{UU}\\
\end{array}
\right)
~,\quad
\mtrx{N} =  \left(
\begin{array}{ccc}
N_{II} & N_{IQ} & N_{IU}\\
N_{QI} & N_{QQ} & N_{QU}\\
N_{UI} & N_{UQ} & N_{UU}\\
\end{array}
\right)\;.
\end{equation}
Finally, $\Theta_\mathrm{C}$ is the set of standard cosmological parameters,
$\Theta_\mathrm{T}$ is the set of topological parameters (e.g., the size, $L$, of the fundamental domain), $\varphi$ is the
orientation of the topology (e.g., the Euler angles), and
$A$ is a single amplitude, scaling the signal covariance matrix (this is equivalent to an overall amplitude in front of the power spectrum in the isotropic case).
Working in pixel space allows for the straightforward application of an arbitrary
mask, including separate masks for intensity and polarization parts of the data.
The masking procedure can also be used to limit the analysis to intensity or polarization only.

Since $\mtrx{C}+\mtrx{N}$ in pixel space is generally poorly conditioned,
we again (following the 2013 procedure) project the data vector and covariance
matrices onto a limited set of orthonormal basis vectors,
select $N_\mathrm{m}$ such modes for comparison, and consider the likelihood marginalized over the remainder of the modes,
\begin{eqnarray}
\lefteqn{p( \fitdata | \mtrx{C}[ \Theta_\mathrm{C},\Theta_\mathrm{T},T],\varphi, A) \propto} \nonumber \\
&&\frac{1}{\sqrt{|A \mtrx{ C} + \mtrx{ N} |_M}}
\exp\left[ - \frac{1}{2} \sum_{n=1}^{N_\mathrm{m}} d_n^* (A \mtrx{ C} + \mtrx{ N})^{-1}_{nn^\prime} d_{n^\prime} \right] \, ,
\label{eq:L_modes}
\end{eqnarray}
where $\mtrx{C}$ and $\mtrx{N}$ are restricted to the
$N_\mathrm{m} \times N_\mathrm{m}$ subspace.

The choice of the basis modes and their number $N_m$ used for analysis
is a compromise between robust invertibility of $ \mtrx{ C} + \mtrx{ N} $ and
the amount of information retained. All the models for which likelihoods
are compared must be expanded in the same set of modes. Thus,
in \citet{planck2013-p19} we used the set of eigenmodes
of the cut-sky covariance matrix of the fiducial best-fit simply-connected
universe, $\mtrx{C}_\mathrm{fid}$, as the analysis basis,
limiting ourselves to the $N_\mathrm{m}$ modes with the largest eigenvalues. For comparison with the numbers we use, a full-sky temperature map with maximum multipole $\ell_\mathrm{max}$ has $(\ell_\mathrm{max}+1)^2-4$ independent modes (four are removed to account for the unobserved monopole and dipole).

The addition of polarization data, with much lower signal-to-noise than the temperature, raises a new question: how is the temperature
and polarization data mix reflected in the limited basis set we project onto?
The most natural choice is the set of eigenmodes of the signal-to-noise matrix
$\mtrx{C}_\mathrm{fid} \mtrx{N}^{-1}$ for the fiducial model, and a restriction of
the mode set based on signal-to-noise eigenvalues \citep[see, e.g.,][]{bjk98}. This, however, requires robust
invertibility of the noise covariance matrix, which, again, is generally
not the case for the smoothed data. Moreover, such a ranking by $S/N$ would inevitably favour the temperature data, and we wish to explore the effect of including polarization data on an equal footing with temperature. We therefore continue to use the eigenmodes of the cut-sky fiducial covariance matrix as our basis. By default, we select the first $N_m$ = 1085 eigenmodes (corresponding to $\ell_\mathrm{max}=32$), though we vary the mode count where it is informative to do so.

\begin{figure}[htbp]
    \centering
        \includegraphics[width=1.0\columnwidth]{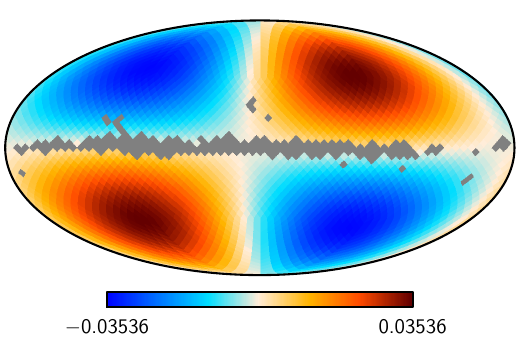}
        \includegraphics[width=1.0\columnwidth]{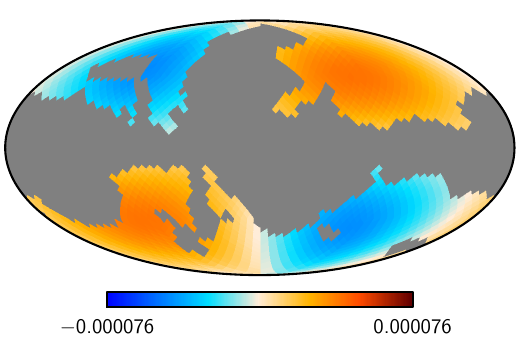}
        \includegraphics[width=1.0\columnwidth]{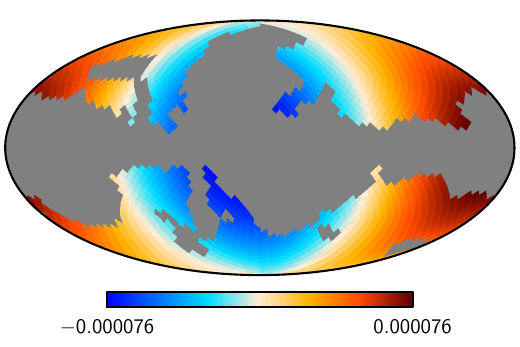}
    \caption{Mode structure plotted as maps for the eigenvector corresponding to the highest-signal eigenvalue of the fiducial simply-connected model. The top map corresponds to temperature, middle to $Q$ polarization, and bottom to $U$ polarization. Masked pixels are plotted in grey.}
    \label{fig:modemaps_highS}
\end{figure}

\begin{figure}[htbp]
    \centering
        \includegraphics[width=1.0\columnwidth]{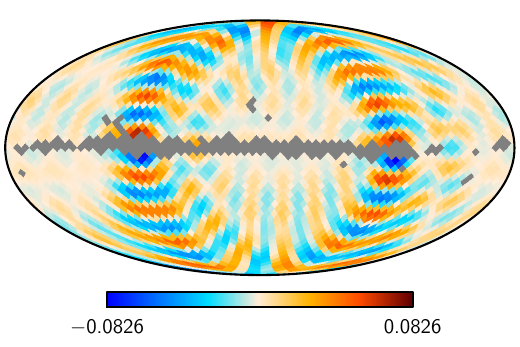}
        \includegraphics[width=1.0\columnwidth]{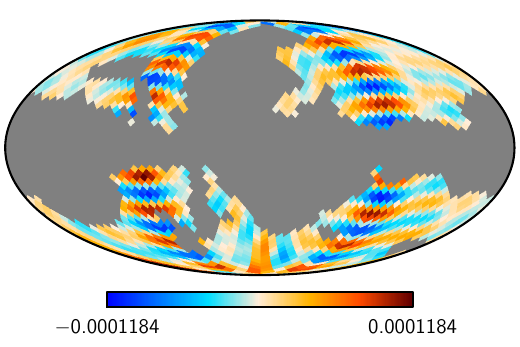}
        \includegraphics[width=1.0\columnwidth]{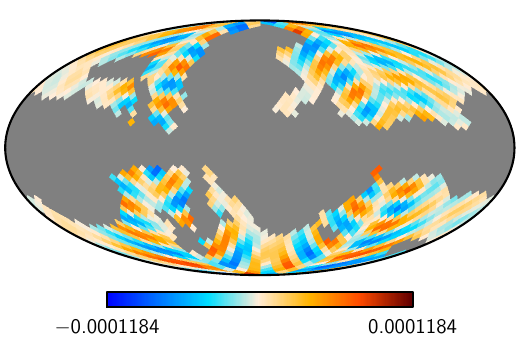}
    \caption{Mode structure plotted as maps for the eigenvector corresponding to the 301st-highest-signal eigenvalue of the fiducial simply-connected model. The top map corresponds to temperature, middle to $Q$ polarization, and bottom to $U$ polarization. Masked pixels are plotted in grey.}
    \label{fig:modemaps_lowS}
\end{figure}

In Fig.~\ref{fig:modemaps_highS} we show $I$, $Q$, and $U$ maps of the highest-eigenvalue (i.e., highest  contribution to the signal covariance) mode for our fiducial simply-connected model. Note that the scale is different for temperature compared to the two polarization maps: the temperature contribution to the mode is much greater than that of either polarization component. We show modes for the masked sky, although in fact the structure at large scales is similar to the full-sky case, rotated and adjusted somewhat to account for the mask. In Fig.~\ref{fig:modemaps_lowS} we show the structure of mode 301, with much lower signal amplitude \rev{(this particular mode was selected at random to indicate the relative ratios of temperature, polarization, and noise)}. Temperature remains dominant, although polarization begins to have a larger effect. Note that at this level of signal amplitude, the pattern is aligned with the mask, and shows a strong correlation between temperature and polarization.

\subsubsection{Evaluating the topological likelihood} 
\label{subs:topology_marge}

The aim of the topological likelihood analysis is to calculate the likelihood as a function of the parameters pertaining to a particular topology, $p(\fitdata | \Theta_\mathrm{T}, T)$. To do so, we must marginalize over the other parameters appearing in Eq.~\eqref{eq:L_modes}, namely $\Theta_\mathrm{C}$, $\varphi$, and $A$, as
\begin{eqnarray}
\lefteqn{p(\fitdata | \Theta_\mathrm{T}, T) =} \nonumber \\ && \int {\rm d} \Theta_\mathrm{C} \, {\rm d} \varphi \, {\rm d}A \, p( \fitdata | \mtrx{C}[ \Theta_\mathrm{C}, \Theta_\mathrm{T},T],\varphi, A) \, p(\Theta_\mathrm{C},\varphi, A).
\label{eq:L_marg_full}
\end{eqnarray}
The complexity of the topological covariance matrix calculation precludes a joint examination of the full cosmological and topological parameter spaces. Instead, we adopt the delta-function prior $p(\Theta_\mathrm{C}) = \delta(\Theta_\mathrm{C} - \Theta_\mathrm{C}^\star)$ to fix the cosmological parameters at their fiducial values, $\Theta_\mathrm{C}^\star$ (as defined in Sect.~\ref{subsub:corr_mat_calc}), and evaluate the likelihood on a grid of topological parameters using a restricted set of pre-calculated covariance matrices. We note that, as discussed in \citet{planck2013-p19}, the ability to detect or rule out a multiply connected topology is insensitive to the values of the cosmological parameters adopted for the calculation of the covariance matrices.

In the setting described above, Eq.~\eqref{eq:L_marg_full} simplifies to
\begin{equation}
p(\fitdata | \Theta_{\mathrm{T}, i}, T) = \int {\rm d} \varphi \, {\rm d}A \, p( \fitdata | \mtrx{C}[ \Theta_\mathrm{C}^\star,\Theta_{\mathrm{T}, i},T],\varphi, A) \, p(\varphi, A),
\label{eq:L_marg}
\end{equation}
where the likelihood at each gridpoint in topological parameter space, $\Theta_{\mathrm{T}, i}$, is equal to the probability of obtaining the data given fixed cosmological and topological parameters and a compactification (i.e., fundamental domain shape and size), marginalized over orientation and amplitude. The calculation therefore reduces to evaluating the Bayesian evidence for a set of gridded topologies. As we focus on cubic torus and slab topologies in this work, we note that the sole topological parameter of interest is the size of the fundamental domain, $L$.

Even after fixing the cosmological parameters, calculating the Bayesian evidence is a time-consuming process, and is further complicated by the multimodal likelihood functions typical in non-trivial topologies. We therefore approach the problem on two fronts. We first approximate the likelihood function using a ``profile likelihood'' approach, as presented in \citet{planck2013-p19}, in which the marginalization in Eq.~\eqref{eq:L_marg} is replaced with maximization in
the four-dimensional space of orientation and amplitude parameters.
Specifically, we maximize the likelihood over the three angles defining the orientation of
the fundamental domain using a three-dimensional Amoeba search \citep[e.g.,][]{Press:1992:NRC:148286},
where at each orientation the likelihood is separately maximized over
the amplitude. Due to the complex structure of the likelihood
surface in orientation space, we repeat this procedure five times with different
starting orientations. \rev{This number of repetitions was chosen as a compromise between computational efficiency and assurance of statistical robustness, after testing of various strategies for the number of repetitions and the distribution of starting points, along with explicit extra runs to test outliers.} To ensure uniform and non-degenerate coverage,
the orientation space is traversed in a cartesian projection of the northern
hemisphere of the three-sphere $S3$ representation of rotations.

The profile likelihood calculation allows rapid evaluation of the likelihood and testing of different models compared with a variety of data and simulations, but it is difficult to interpret in a Bayesian setting. As we show below, however, the numerical results of profiling over this limited set of parameters agree numerically very well with the statistically correct marginalization procedure.

Our second approach explicitly calculates the marginalized likelihood,  Eq.~\eqref{eq:L_marg}, allowing full Bayesian inference at the cost of increased computation time. We use the public \multinest\footnote{\url{http://www.mrao.cam.ac.uk/software/multinest/}}
code \citep{feroz:multinest1,feroz:multinest2,feroz:multinest3}---optimized for exploring multimodal probability distributions in tens of dimensions---to compute the desired evidence values via nested sampling \citep{skilling:2004}. \multinest\ is run in its importance nested sampling mode~\citep{feroz:multinest3} using 200 live points, with tolerance and efficiency set to their recommended values of 0.5 and 0.3, respectively. The final ingredient needed to calculate the evidence values are priors for the marginalized parameters. We use a log prior for the amplitude, truncated to the range $0.1 \le A \le 10$, and the Euler angles are defined to be uniform in $0 \le \alpha < 2\pi$, $-1 \le \cos\beta \le 1$, and $0 \le \gamma < 2\pi$, respectively; \multinest\ is able to wrap the priors on $\alpha$ and $\gamma$. The combined code will be made public as part of the \anicosmo\footnote{\url{http://www.jasonmcewen.org/}} package \citep{mcewen:bianchi}.

It is worth noting that this formalism can be extended to compare models with different compactifications (or the simply connected model) using Bayesian model selection: the only additional requirements are priors for the topological parameters. Taking the current slab and cubic torus topologies as examples, by defining a prior on the size of the fundamental domain one can calculate the evidence for each model. Assuming each topology is equally likely a~priori, i.e., that $p(T_{\rm slab}) = p(T_{\rm cub})$, one can then write down the relative probability of the two topologies given the data:
\begin{equation}
\frac{ p(T_{\rm slab} | \fitdata) }{ p(T_{\rm cub} | \fitdata) } \simeq \frac{ \sum_i p(L_i | T_{\rm slab}) \, p(\fitdata | L_i, T_{\rm slab}) }{ \sum_j p(L_j^\prime | T_{\rm cub}) \, p(\fitdata | L_j^\prime, T_{\rm cub}) }.
\label{eq:top_model_select}
\end{equation}
Unfortunately, it is difficult to provide a physically-motivated proper prior distribution for the size of the fundamental domain. Even pleading ignorance and choosing a ``na{\"\i}ve'' uniform prior would require an arbitrary upper limit to $L$ whose exact value would strongly influence the final conclusion. For this reason, we refrain from extending the formalism to model selection within this manuscript.

\subsubsection{Bianchi models} 
\label{subs:bianchi}


While physically the cosmological densities describing Bianchi models
should be identified with their standard $\Lambda$CDM counterparts, in
previous analyses unphysical models have been considered in which the
densities are allowed to differ. The first coherent analysis of
\bianchiviih\ models was performed by \citet{mcewen:bianchi}, where the
$\Lambda$CDM and Bianchi densities are coupled and all cosmological
and Bianchi parameters are fit simultaneously.
In the analysis of \citet{planck2013-p19}, in order to compare with all
prior studies both coupled and decoupled models were analysed. We
consider the same two models here: namely, the physical {\em
  open-coupled-Bianchi\/} model where an open cosmology is considered
(for consistency with the open \bianchiviih\ models), in which the Bianchi
densities are coupled to their standard cosmological counterparts; and
the phenomenological {\em
  flat-decoupled-Bianchi\/} model where a flat cosmology is considered
and in which the Bianchi densities are decoupled.

We firstly carry out a full Bayesian analysis for these two
\bianchiviih\ models, repeating the analysis performed in
\citet{planck2013-p19} with updated \Planck\ temperature data. The
methodology is described in detail in \citet{mcewen:bianchi} and
summarized in \citet{planck2013-p19}. The complete posterior
distribution of all Bianchi and cosmological parameters is sampled and
Bayesian evidence values are computed to compare \bianchiviih\ models
to their concordance counterparts. Bianchi temperature signatures are
simulated using the {\tt
  Bianchi2}\footnote{\url{http://www.jasonmcewen.org/}} code
\citep{mcewen:bianchi}, while the \anicosmo\ code
is used to perform the analysis, which in turn
uses \multinest\
to sample the posterior distribution and compute evidence values.

To connect with polarization data, we secondly analyse polarization
templates computed using the best-fit parameters from the analysis of
temperature data. For the resulting small set of best-fit models,
polarization templates are computed using the approach of
\citet{pontzen:2007} and \citet{pontzen:2009}, and have been provided by
\citet{PontzenPrivate}.
These \bianchiviih\ simulations are
more accurate than those considered for the temperature analyses performed
here and in previous works (see, e.g., \citealt{planck2013-p19},
\citealt{mcewen:bianchi}, \citealt{bridges:2008},
\citealt{bridges:2006b}, \citealt{jaffe:2005, jaffe:2006c,
  jaffe:2006a, jaffe:2006b}), since the
recombination history is modelled.
\rev{The overall morphology of the patterns are consistent between the codes;}
the largest effect of incorporating
the recombination history is its impact on the polarization fraction,
although the amplitude of the temperature component can also vary by
approximately 5\,\% (which is calibrated in the current analysis, as
described below).

Using the simulated \bianchiviih\ polarization templates computed
following \citet{pontzen:2007} and \citet{pontzen:2009}, and provided by \citet{PontzenPrivate},
we perform a maximum-likelihood fit
for the amplitude of these templates using \Planck\ polarization data
(a full Bayesian evidence calculation of the complete temperature and
polarization data set incorporating the more accurate Bianchi models
of \citealt{pontzen:2007} and \citealt{pontzen:2009} is left to future
work).
The likelihood in the Bianchi scenario is identical to that considered in
\citet{planck2013-p19} and \citet{mcewen:bianchi}; however, we now consider the
Bianchi and cosmological parameters fixed and simply introduce a
scaling of the Bianchi template.  The resulting likelihood reads:
\begin{equation}
P( \vec{d} \ \vert \ \lambda, \vec{t})
\propto \exp \biggl[ -\frac{1}{2}
(\vec{d} - \lambda \vec{t})^\dagger
(\mtrx{C}+\mtrx{N})^{-1}
(\vec{d} - \lambda \vec{t}) \biggr] ,
\end{equation}
where $\vec{d}$ denotes the data vector, $\vect{t} =
\vect{b}(\bparam^\star)$ is the Bianchi template for best-fit Bianchi
parameters $\bparam^\star$, $\mtrx{C} =\mtrx{C(\cosmoparam)}$ is the
cosmological covariance matrix for the best-fit cosmological
parameters $\cosmoparam^\star$, $\mtrx{N}$ is the noise covariance, and
$\lambda$ is the introduced scaling parameter (the effective vorticity
of the scaled Bianchi component is simply $\lambda (\omega/H)_0$).

In order to effectively handle noise and partial sky coverage the data
are analysed in pixel space. We restrict to polarization data only here
since temperature data are used to determine the best-fit Bianchi
parameters. The data and template vectors thus contain unmasked $Q$
and $U$ Stokes components only and, correspondingly, the cosmological
and noise covariance matrices are given by the polarization ($Q$ and $U$) subspace of Eq.~\eqref{eq:polcorrmat},
and again contain unmasked pixels only.

The maximum-likelihood (ML) estimate of the template amplitude is
given by $\lambda^{\rm ML} = \vec{t}^\dagger (\mtrx{C}+\mtrx{N})^{-1} \vec{d} /
\left[\vec{t}^\dagger(\mtrx{C}+\mtrx{N})^{-1} \vec{t}\right]$
and its dispersion by $\Delta
\lambda^{\rm ML} = [ \vec{t}^\dagger (\mtrx{C}+\mtrx{N})^{-1} \vec{t} ]^{-1/2}$
\citep[see, e.g.,][]{kogut:1997, jaffe:2005}.  If \Planck\ polarization data
support the best-fit Bianchi model found from the analysis of
temperature data we would expect $\lambda^{\rm ML} \simeq 1$.  A
statistically significant deviation from unity in the fitted amplitude
can thus be used to rule out the Bianchi model using polarization
data.

As highlighted above, different methods are used to simulate Bianchi
temperature and polarization components, where the amplitude of the
temperature component may vary by a few percent between methods. We
calibrate out this amplitude mismatch by scaling the polarization
components by a multiplicative factor fitted so that the temperature
components simulated by the two methods match, using a
maximum-likelihood template fit again, as described above.



\subsection{Simulations and validation} 
\label{sub:Simulations}

\subsubsection{Topology} 
\label{sec:applic_simulations}

\begin{figure}
\centering
   \includegraphics[width=1.0\columnwidth]{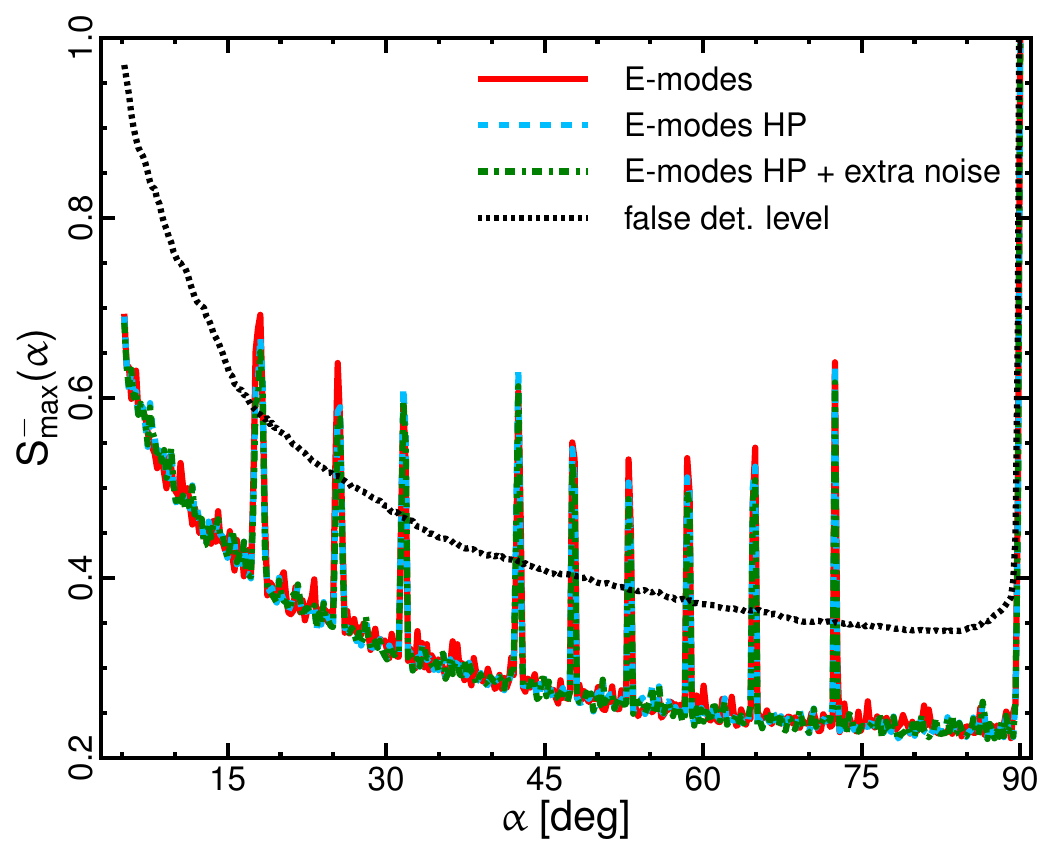}
\caption{An example of the $S_{\rm max}^{-}$ statistic as a function
  of circle radius $\alpha$ for a simulated CMB $E$-mode map of a
  universe with the topology of a cubic 3-torus with dimensions $L =
  2H_0^{-1}$. To this map noise was added corresponding to the \smica\ map.
  The thick overlapping curves show the statistic for
  simulated polarization maps with angular resolution and noise level
  corresponding to the \Planck\ \smica\ map for three cases: without high-pass filtering (solid, red); with filtering (dashed, blue); and with a 5\,\% larger
  amplitude of noise (dot-dashed, green).
  The dotted line shows the false detection level established such that
fewer than 1\,\% of 300 Monte Carlo simulations of the high-pass filtered \Planck\ \smica\ polarization maps, smoothed and masked in the same way as the
data, would yield a false event.}
\label{fig:smax_t222}
\end{figure}

\paragraph{Matched circles.} 
\label{par:circles_validation}
Before beginning the search for pairs of matched circles in the
\Planck\ data, we validate our algorithm using the same simulations as
employed for the \citet{planck2013-p19} and \citet{bielewicz2012} papers, i.e., the
CMB sky for a universe with a three-torus topology for which the
dimension of the cubic fundamental domain is $L=2\,H_0^{-1}$, well within the last-scattering surface.
We computed the $a_{\ell m}$ coefficients up to the multipole of order
$\ell=500$ and convolved them with the same smoothing beam profile as used for the
\Planck\ \smica\ map. To the map was added noise corresponding to the
\smica\ map. In particular, we verified that our code is able to find
all pairs of matched circles in such a map. The statistic $S_{\rm
  max}^{-}(\alpha)$ for the $E$-mode map is shown in Fig.~\ref{fig:smax_t222}.

Because for the baseline analysis we use high-pass filtered maps, we
also show the analysis of the \smica\ $E$-mode map high-pass filtered
so that the lowest order multipoles ($\ell < 20$) are removed from the
map (the multipoles in the range $20 \le \ell \le 40$ are apodized between 0 and 1 using a
cosine as defined in \citealt{planck2014-a11}). The high-pass filtering does not decrease our ability to
detect a multiply-connected topology using the matched-circle
method. This is consistent with the negligible sensitivity of the
matched-circle statistic to the reionization signal,
studied by \citet{bielewicz2012}. This is a consequence of the
weighting of the polarization data by a factor proportional to $\ell^2$
employed in the transformation from the Stokes parameters $Q$ and $U$ to
an $E$-mode map, which effectively filters out the largest-scale multipoles from the
data. This test shows that the matched-circle method, contrary to the likelihood method,
predominantly exploits the topological signal in the CMB
anisotropies at moderate angular scales.

We also checked robustness of detection with respect to noise
level in order to account for small discrepancies between the noise level in the \Planck\ FFP8 simulations and the 2015 data \citep{planck2014-a14}.
We repeated the analysis for the high-pass filtered map with added noise with 5\,\% larger
amplitude than for the original map. As we can see in \fig\ref{fig:smax_t222},
the statistic changes negligibly.

The intersection of the peaks in the matching statistic with the false
detection level estimated for the CMB map corresponding to the
simply-connected universe defines the minimum radius of the correlated circles that
can be detected for this map. \rev{We estimate the minimum radius by extrapolating the height of the peak
with radius $18^\circ$ seen in Fig.~\ref{fig:smax_t222} towards
smaller radii. This allows for a rough estimation of the radius, with a precision of a few degrees. However, better precision is not required, because for small minimum radius (as obtains here) constraints on the size of the fundamental domain are not very sensitive to differences of the minimum radius of order a few degrees. As we can see in Fig.~\ref{fig:smax_t222}, the minimum radius $\alpha_{\rm min}$ takes a value in the range from $10^\circ$ to around $15^\circ$. To be conservative we take the upper end of this range for computation of the constraints on the size of the fundamental domain, and thus use $\alpha_{\rm min} \simeq 15^\circ$.}

\paragraph{Likelihood.} 
\label{par:likelihood_validation}
To validate and compare the performance of the two likelihood methods, we perform two sets of tests: a null test using a simulation of a simply connected universe, and a signal test using a simulation of a toroidal universe with $L = 4H_0^{-1}$. The two test maps are generated at $N_{\rm side} = 16$ and are band-limited using a $640\arcm$ Gaussian beam. Diagonal (white) noise is added with pixel variances $\sigma_I^2 = 0.04\,\mu{\rm K}^2$ and $\sigma_{Q/U}^2 = 0.16\,\mu{\rm K}^2$, comparable to the expected eventual level of \Planck's 143\GHz\ channel. For clarity of interpretation, no mask is used in these tests; in this setting, the eigenmodes of the fiducial covariance matrix are linear combinations of the spherical harmonics at fixed wavenumber $\ell$.
As fully exploring the likelihood is much more time-consuming than profiling it, we generate a complete set of test results---analyses of the two test maps using cubic torus and slab covariance matrices on a fine grid of fundamental domain scales---using the profile-likelihood code, and aim to verify the main cubic torus results using the marginalized likelihood generated with \anicosmo. Note that to speed up the calculation of the marginalized likelihood we use a slightly smaller band-limit ($\elmax = 32$) than in the profile-likelihood calculation ($\elmax = 40$); with our choice of smoothing scale and mode count, and considering the full sky for validation purposes, we obtain the same eigenbasis (and therefore analyse the same projected data) in both cases.

\begin{figure}[htbp]
    \centering
        \includegraphics[width=1.0\columnwidth]{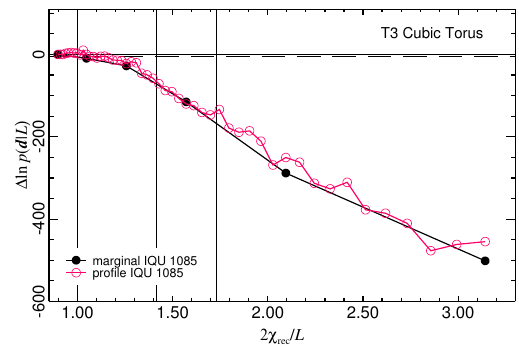}
    \caption{The likelihood function for the fundamental domain scale of a cubic torus derived from simulations of a simply-connected universe, calculated through marginalization (black, filled circles) and profiling (pink, empty circles). The horizontal axis gives the inverse of the length of a side of the fundamental domain, relative to the distance to the last-scattering surface. The vertical lines mark the positions where $\chi_{\rm rec}$ is equal to various characteristic sizes of the fundamental domain, namely the radius of the largest sphere that can be inscribed in the domain, $\mathcal{R}_{\rm i} = L/2$, the smallest sphere in which the domain can be inscribed, $\mathcal{R}_{\rm u} = \sqrt{3}L/2$, and the intermediate scale $\mathcal{R}_{\rm m} = \sqrt{2}L/2$.}
\label{fig:figures_null_test_torus_L_posterior}
\end{figure}

\begin{figure}[htbp]
    \centering
        \includegraphics[width=1.0\columnwidth]{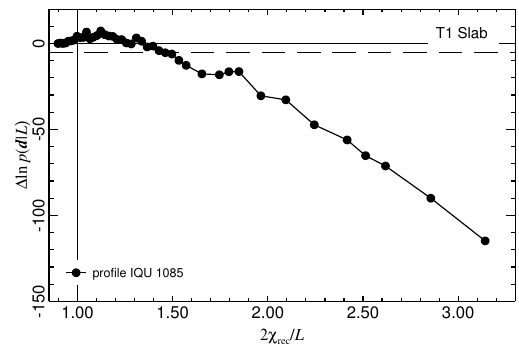}
    \caption{The profile likelihood function for the fundamental domain scale of a slab topology derived from simulations of a simply-connected universe. The vertical line marks the position where $\chi_{\rm rec}$ is equal to the radius of the largest sphere that can be inscribed in the domain, $\mathcal{R}_{\rm i} = L/2$ (for slab spaces, the other two characteristic sizes are infinite).}
\label{fig:figures_null_test_slab_L_posterior}
\end{figure}

The results for the null test---in the form of the likelihood function for the fundamental domain scale of the assumed topology---are plotted in Fig.~\ref{fig:figures_null_test_torus_L_posterior} for cubic tori and Fig.~\ref{fig:figures_null_test_slab_L_posterior} for slabs. Concentrating initially on the cubic tori, we see that the likelihoods derived from the two codes agree. In both cases, the likelihood is found to be maximal for fundamental domain scales larger than the horizon, and the small-$L$ cubic tori are very strongly disfavoured: $p({\fitdata | L=7H_0^{-1}}) / p({\fitdata | L=2H_0^{-1}}) \sim 10^{217}$. Note that the \anicosmo\ likelihood curve contains errors on the likelihood at each $L$ considered, but these are orders of magnitude smaller than the changes in likelihood between points (typical \multinest\ uncertainties yield errors of order $0.1$ in log-likelihood).

In both cases, the profile likelihood exhibits a mild rise around the horizon scale, due to chance alignments along the matched faces of the fundamental domain. It is slightly stronger in the slab case since the probability of such alignments is greater with only a single pair of faces.

\begin{figure}[htbp]
    \centering
        \includegraphics[width=1.0\columnwidth]{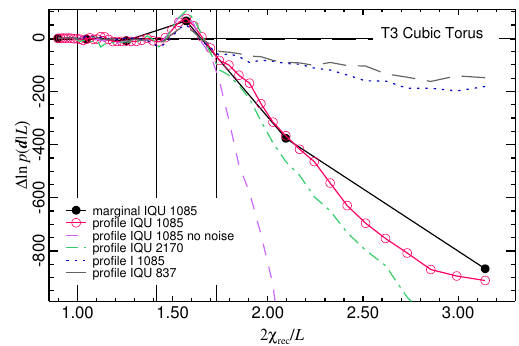}
    \caption{The likelihood function for the fundamental domain scale of a cubic torus derived from a simulation of a toroidal universe with $L=4H_0^{-1}$ or $2\chi_\mathrm{rec}/L=1.5$. The results from the profile-likelihood analysis (pink, clear circles) closely match those from the full marginalized likelihood (black, filled circles). Overlaid are additional profile likelihoods demonstrating the effects of changing the mode count and composition. In order of increasing constraining power, they utilize 837 $IQU$ modes (grey, long dashed), 1085 $I$ modes (blue, dotted), 1085 $IQU$ modes (pink, clear circles), 2170 $IQU$ modes (green, dot-dashed), and finally 1085 noiseless $IQU$ modes (purple, dashed). Adding low-$\ell$ (ideally low-noise) polarization greatly increases the constraining power of the data.}
\label{fig:figures_torus_test_torus_L_posterior}
\end{figure}

\begin{figure}[htbp]
    \centering
        \includegraphics[width=1.0\columnwidth]{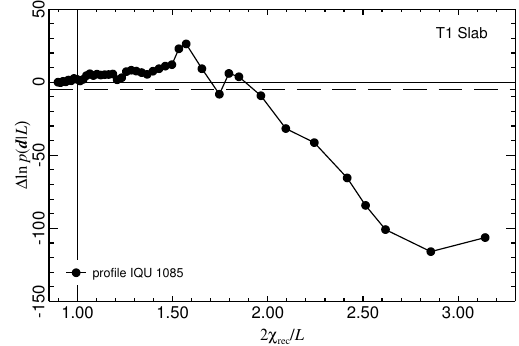}
    \caption{The profile likelihood function for the fundamental domain scale of a slab derived from a simulation of a toroidal universe with $L=4H_0^{-1}$ or $2\chi_\mathrm{rec}/L=1.5$.}
\label{fig:figures_torus_test_slab_L_posterior}
\end{figure}

The results for the tests on the toroidal simulation are shown in Fig.~\ref{fig:figures_torus_test_torus_L_posterior} for toroidal covariance matrices and Fig.~\ref{fig:figures_torus_test_slab_L_posterior}
for slab covariance matrices. Concentrating first on the results employing toroidal covariance matrices, the correct fundamental domain scale is clearly picked out by both the profile and full likelihood codes, with the simply connected case strongly disfavoured at a likelihood ratio of $p({\fitdata | L=4H_0^{-1}}) / p({\fitdata | L=7H_0^{-1}}) \sim 10^{28}$.
Turning to the results derived using slab covariance matrices, we see that---as expected from the Kullback-Leibler divergence analysis of Sect.~\ref{sec:KL}---the correct fundamental domain scale is also found using the slab profile likelihood. Although, as also expected, the peak is not quite as pronounced when using the wrong covariance matrix, the simply connected universe is still overwhelmingly disfavoured at a ratio of $p({\fitdata | L=4H_0^{-1}}) / p({\fitdata | L=7H_0^{-1}}) \sim 10^{11}$.

The speed of the profile-likelihood analysis allows for the effects of changing the mode count, composition, and noise level to be investigated.
We repeat the toroidal test using intensity-only ($I$) and full ($IQU$) covariance matrices, retaining between 837 and 2170 modes at a time. For the smoothing scale employed in our tests, the 838th mode is the first to be dominated by polarization; runs using up to 837 $IQU$ modes are therefore dominated by intensity information.
The results of this investigation are contained in Fig.~\ref{fig:figures_torus_test_torus_L_posterior}. The most striking conclusion is that the impact of adding temperature modes to the analysis is dwarfed by the impact of adding low-$\ell$ polarization information,
even though the temperature modes are effectively noiseless. This conclusion is supported by the observation from Fig.~\ref{fig:topo_kl_lmax} that the Kullback-Leibler divergence grows most rapidly at low $\ell$.



\subsubsection{Bianchi} 
\label{ssub:bianchi}

The Bayesian analysis of \bianchiviih\ models using temperature data
is performed using the \anicosmo\ code,
which has been extensively validated by \citet{mcewen:bianchi},
and was used to perform the Bianchi analysis of \citet{planck2013-p19}.
The maximum-likelihood template fitting method used to analyse
polarization data is straightforward and has been validated on
simulations, correctly recovering the amplitude of templates artificially
embedded in simulated CMB observations.




\section{Results} 
\label{sec:results}

\subsection{Topology} 
\label{sub:results_topology}

\subsubsection{Matched circles} 
\label{ssub:matched_circles}

\begin{figure}
\centering
   \includegraphics[width=1.0\columnwidth]{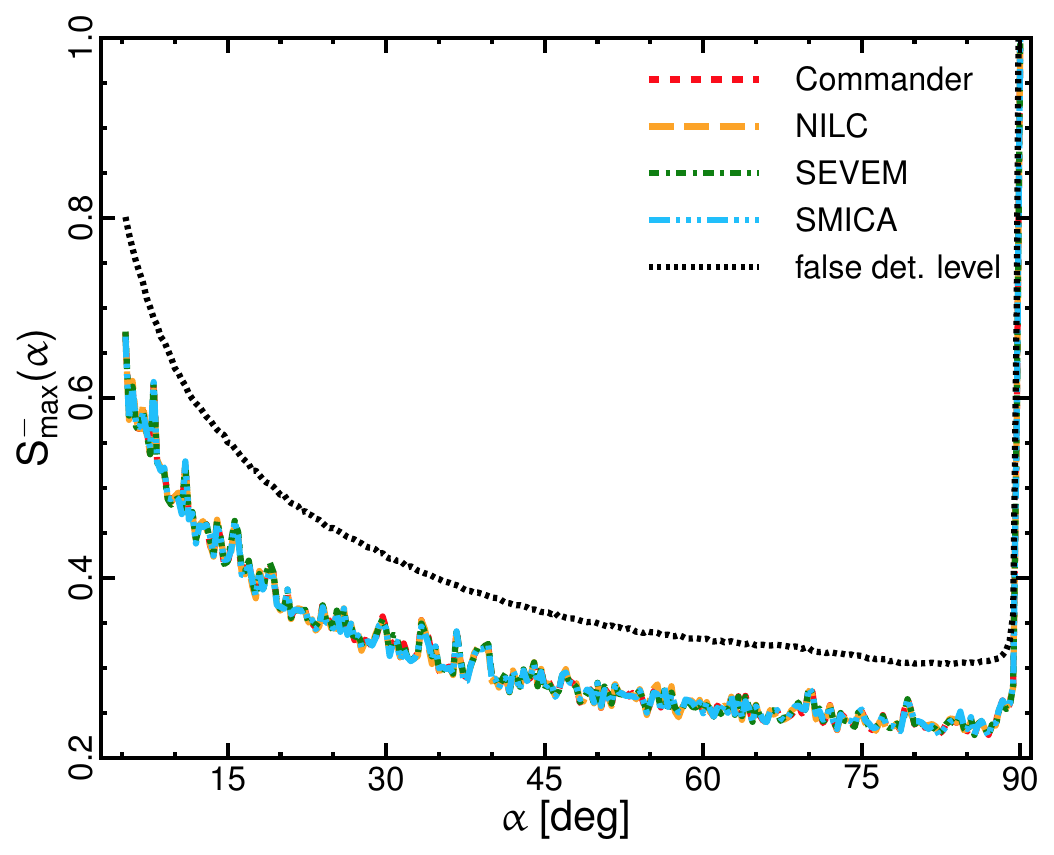}
   \includegraphics[width=1.0\columnwidth]{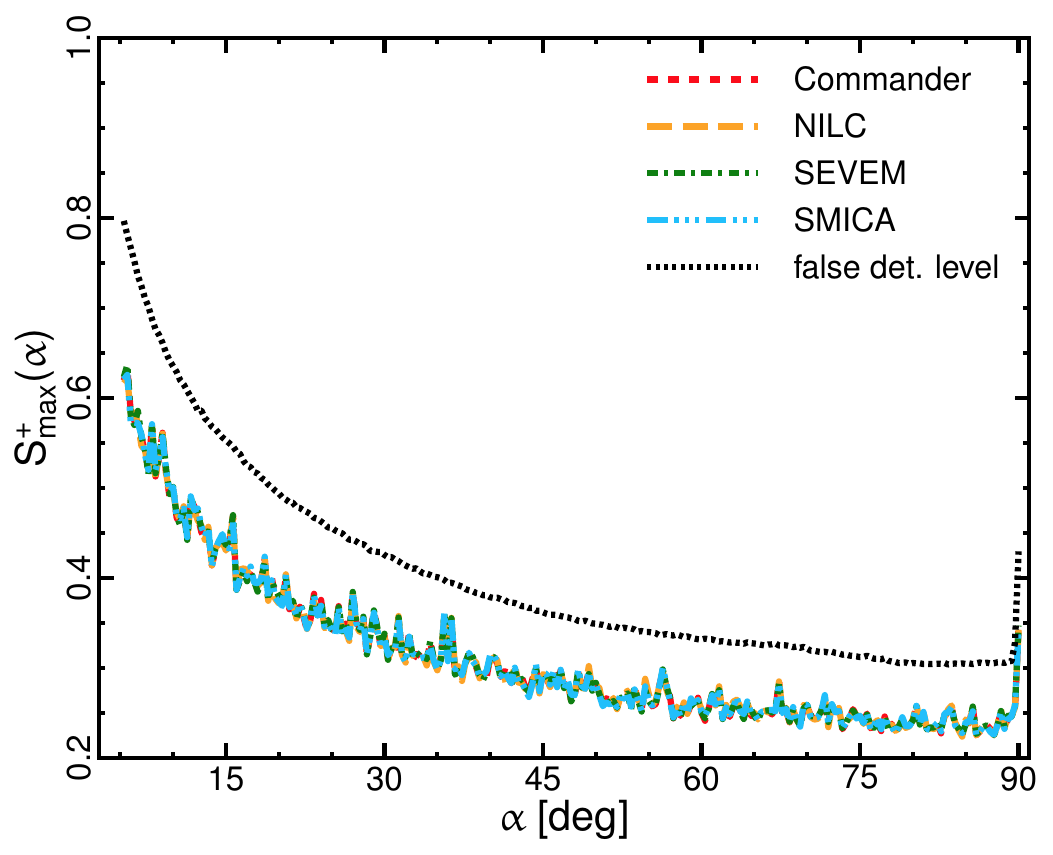}
\caption{$S_{\rm max}^-$ (upper) and $S_{\rm max}^+$ (lower)
  statistics as a function of circle radius $\alpha$ for the
  \Planck\ \commander\ (short-dashed red line), \nilc\ (long dashed
  orange line), \sevem\ (dot-dashed green line), and \smica\ (three
  dot-dashed blue line) 2015 temperature maps. The dotted line shows the false detection
  level established such that fewer than 1\,\% of 300 Monte Carlo
  simulations of the \smica\ CMB temperature map, smoothed
  and masked in the same way as the data, would yield a false event. The peak
  at $90^\circ$ corresponds to a match between two copies of the same
  circle of radius $90^\circ$ centred around two antipodal points.
  }
\label{fig:smax_temp}
\end{figure}
\begin{figure}
\centering
   \includegraphics[width=1.0\columnwidth]{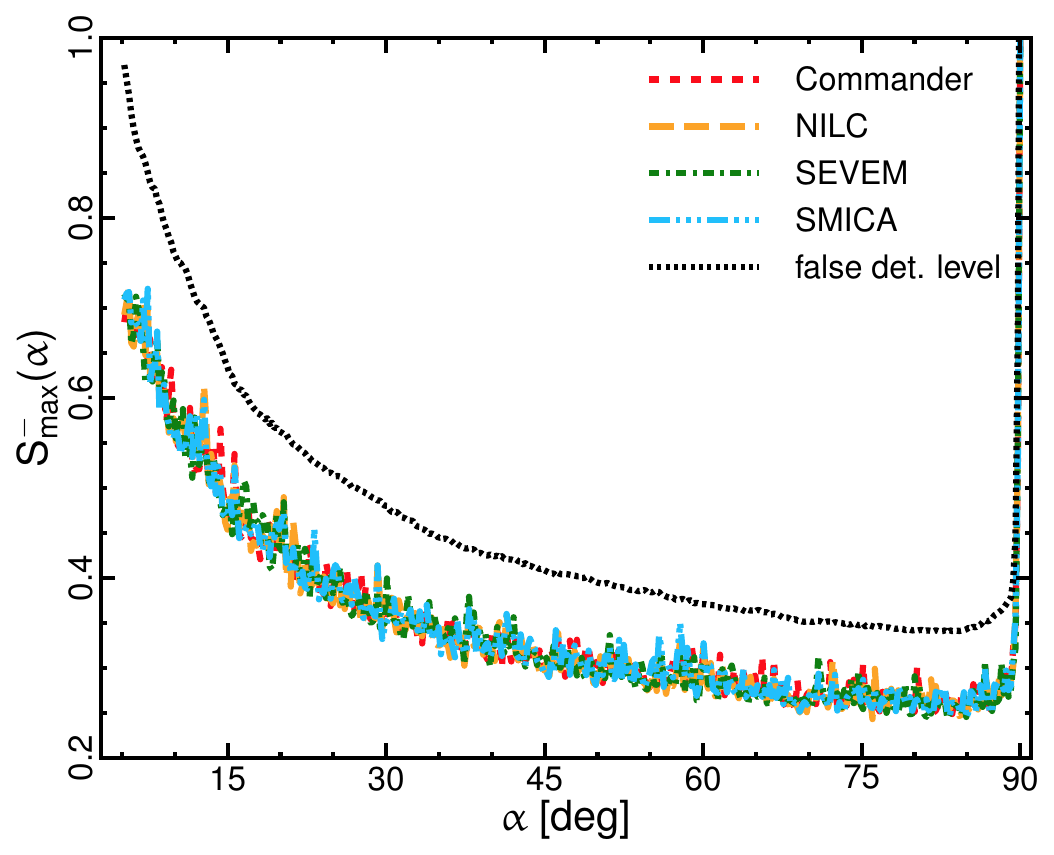}
   \includegraphics[width=1.0\columnwidth]{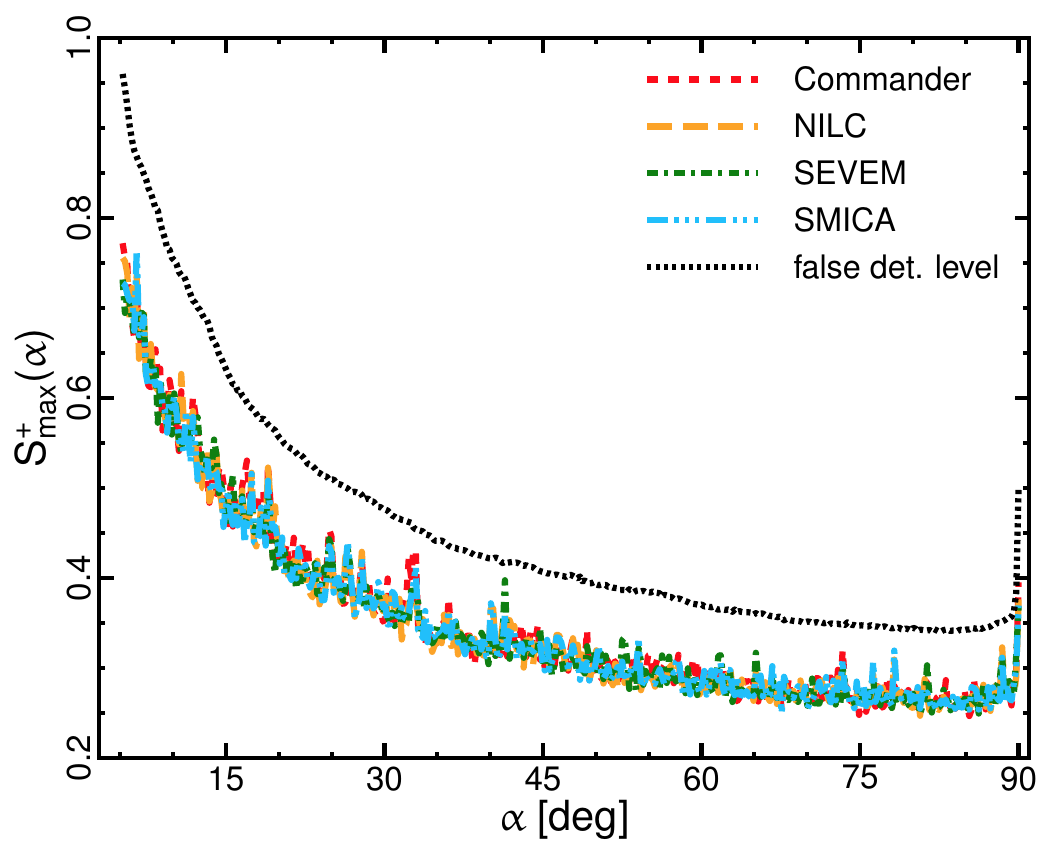}
\caption{$S_{\rm max}^-$ (upper) and $S_{\rm max}^+$ (lower)
  statistics as a function of circle radius $\alpha$ for the
  \Planck\ \commander\ (short-dashed red line), \nilc\ (long dashed
  orange line), \sevem\ (dot-dashed green line), and \smica\ (three
  dot-dashed blue line) $E$-mode maps. The dotted line shows the false detection
  level established such that fewer than 1\,\% of 300 Monte Carlo
  simulations of the \smica\ CMB $E$-mode map, smoothed
  and masked in the same way as the data, would yield a false event. The peak
  at $90^\circ$ corresponds to a match between two copies of the same
  circle of radius $90^\circ$ centred around two antipodal points.
  }
\label{fig:smax_polar}
\end{figure}

We show the matched-circle statistic for the CMB temperature and
$E$-mode maps in \fig\ref{fig:smax_temp} and~\ref{fig:smax_polar},
respectively. We do not find any statistically significant
correlation of circle pairs in any map. Results for the temperature
maps are consistent with the \citet{planck2013-p19} results. 
\rev{As discussed in Sect.~\ref{par:circles_validation}, }
the minimum radius at which the peaks expected for the matching statistic are
larger than the false detection level for the polarization map is around
$\alpha_{\rm min} \simeq 15^\circ$. Thus, we can
exclude, at the confidence level of 99\,\%, any topology that predicts
matching pairs of back-to-back circles larger than this radius,
assuming that the relative orientation of the fundamental domain and mask
allows its detection. This implies that in a flat universe described
otherwise by the fiducial $\Lambda$CDM model, a 99\,\%
confidence-limit lower bound on the size of the fundamental domain is
${\cal R}_\mathrm{i} = L/2\gtrsim\chi_\mathrm{rec}\cos(\alpha_\mathrm{min}) =
0.97\,\chi_\mathrm{rec}$ or $L\gtrsim6.0H_0^{-1}$.  
This is slightly stronger than the constraint obtained for the
analysis of the 2013 \Planck\ temperature maps,
i.e., $0.94\,\chi_\mathrm{rec}$ \citep{planck2013-p19}. Note that the limits from polarization are at least as strong as those from temperature despite the considerably smaller amount of sky considered in the polarization analysis (40\,\% compared to 74\,\%).

\subsubsection{Likelihood} 
\label{ssub:likelihood}

\begin{figure}[htbp]
    \centering
        \includegraphics[width=1.0\columnwidth]{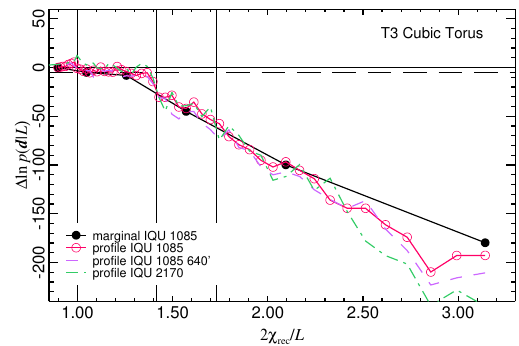}
    \caption{The likelihood for the fundamental domain scale of a cubic torus derived from the \Planck\ low-$\ell$ data set. As in testing, the likelihoods calculated via profiling (pink, clear circles) and marginalization (black, solid circles) agree well. The impact of increasing the polarization content through additional smoothing (purple, dashed) or modes (green, dot dashed) is diminished compared to the test setting due to boosted noise.}
\label{fig:figures_torus_LFI_L_posterior}
\end{figure}

\begin{figure}[htbp]
    \centering
        \includegraphics[width=1.0\columnwidth]{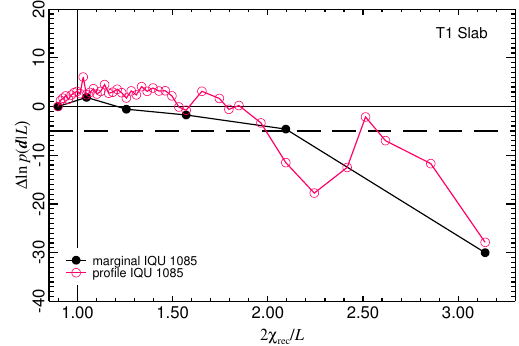}
    \caption{The likelihood for the fundamental domain scale of a slab topology derived from the \Planck\ low-$\ell$ data set  via profiling (pink, clear circles) and marginalization (black, solid circles).}
\label{fig:figures_slab_LFI_L_posterior}
\end{figure}

The results of applying the two likelihood codes to the \Planck\ low-$\ell$ data are plotted in Fig.~\ref{fig:figures_torus_LFI_L_posterior} for cubic tori and Fig.~\ref{fig:figures_slab_LFI_L_posterior} for slabs.
As with the null test, small-$L$ topologies are strongly ruled out, and the likelihoods are maximized at scales
approaching or exceeding the horizon. In the marginalized case, we find that the maximum-likelihood fundamental
domain scale of a cubic torus is $L=7H_0^{-1}$, with scales of $L\le6H_0^{-1}$ disfavoured at greater than
$3.2\,\sigma$; for slabs, we find the likelihood to be peaked at $L=6H_0^{-1}$ (just inside the last-scattering surface),
though $L=7H_0^{-1}$ is allowed at $1.9\,\sigma$.

We have investigated the shape of the likelihood as a function of the slab orientation, and find that it is strongly peaked at an orientation such that the induced matched circles lie partially within the large polarization mask (retaining only 47\,\% of the sky). These orientations therefore do not benefit from the extra discriminatory power of the polarization and its correlation with temperature.

As noted when analysing simulations of simply-connected models (Figs.~\ref{fig:figures_null_test_torus_L_posterior}
--\ref{fig:figures_null_test_slab_L_posterior}), the profile likelihoods also show a mild rise around the horizon
scale for the \Planck\ data: this rise therefore cannot be interpreted as evidence for a multiply connected topology.
We found a similar effect with both profile and marginalized likelihoods in \citet{planck2013-p19} using simulations of
simply connected universes and the \Planck\ temperature data, though this peak was considerably more pronounced for the profile likelihood. Further investigation is required to determine whether this extra-horizon rise is still present in the marginalized likelihood with the present data.

Though the small-$L$ topologies are strongly constrained by the data, we note that the dropoff in likelihood is not as sharp as that observed in our null test (or, indeed, the signal test) presented in Sect.~\ref{par:likelihood_validation}. There are several reasons for this behaviour. The default smoothing scale used in the low-$\ell$ \Planck\ data set ($440\arcm$ Gaussian for intensity, no smoothing for polarization) means all of the first 1085 modes are in fact temperature-dominated. \rev{As our eigenbasis is constructed in decreasing-eigenvalue order, the larger the smoothing scale, the more small-scale temperature modes are damped in comparison to large-scale polarization modes, and thus the earlier polarization modes appear in the basis. Reducing the smoothing scale from $640\arcm$ (as used in testing) to $440\arcm$ (as used in the data) means the first polarization-dominated mode no longer appears in the 1085 highest-eigenvalue modes we take as our fiducial basis.} Exploiting once more the speed of the profile likelihood code, we therefore explore the dependence of the likelihoods on the number of modes and their composition in  Fig.~\ref{fig:figures_torus_LFI_L_posterior}. We introduce polarization-dominated eigenmodes into the analysis in two ways: by simply doubling the number of modes and by retaining the mode count but applying additional smoothing (to bring the effective Gaussian smoothing FWHM to $640\arcm$). Though the constraints on small-$L$ topologies do become tighter when polarization-dominated modes are included, their impact remains weaker than in testing.  This is because the noise in the low-$\ell$ data is significantly higher than in our tests---the typical diagonal covariance matrix element is $\sigma_{IQU}^2 = 4.0\,\mu{\rm K}^2$---and, finally, because we must use a sky cut, which can hide excess correlations. This strongly motivates repeating this analysis with the full multifrequency \Planck\ polarization data when they become available.

Converting these likelihoods into Bayesian constraints on the size of the fundamental domain is not straightforward. Absent a proper prior giving an upper limit on the size of the fundamental domain, there is always infinite parameter volume available for ever-larger fundamental domains. Hence, these likelihood plots should be considered the full summary of the 2015 \Planck\ data for these models. Nonetheless, it is often useful to consider a fall-off in the likelihood of $\Delta\ln{\cal L}<-5$ as roughly equivalent to a
$3\,\sigma$---$99\,\%$ confidence level---limit, the location of which we approximate by interpolating between calculated likelihood points. For the cubic T3 torus, we find that the marginal likelihoods of the combined temperature and polarization \lowTP\ data require that the length of an edge of the fundamental domain satisfies
$L>6 H_0^{-1}$ 
at this significance, or equivalently that the radius of the largest inscribed sphere in the fundamental domain is
${\cal R}_i>0.97\,\chi_\mathrm{rec}$  
(recall that $\chi_\mathrm{rec}\simeq3.1H_0^{-1}$ is the comoving distance to the last-scattering surface.)
The profile likelihood gives the somewhat weaker limit, ${\cal R}_i>0.79\,\chi_\mathrm{rec}$.
 For the T1 slab, we have
$L>3.5H_0^{-1}$ or ${\cal R}_i>0.56\,\chi_\mathrm{rec}$. 
Because the temperature data on the relevant scales are still dominated by the cosmological signal, and the polarization noise remains large, these results are only slightly stronger than those presented in \citet{planck2013-p19}.

\subsection{Bianchi} 
\label{sub:bianchi}


\Planck\ temperature data are masked and analysed for evidence of a
\bianchiviih\ component, using the prior parameter ranges adopted in
\citet{mcewen:bianchi} and \citet{planck2013-p19}. Cleaned temperature
maps for each of the four component separation techniques are
examined, where the mask defined for each technique is applied. The
natural log-Bayes factors for the Bianchi models relative to their standard
cosmological counterparts are shown in
\tbl{\ref{tbl:bianchi_evidences}}. The Bayes factors are broadly
consistent across the component separation methods.  Most Bayes
factors are similar to the values computed in \citet{planck2013-p19},
with the exception of the analysis of the left-handed
flat-decoupled-Bianchi model with \sevem\ data, which has increased,
but which is now more consistent with the other component separation methods.

\begin{table*}
\caption{Natural log-Bayes factors of Bianchi models relative to equivalent
  $\Lambda$CDM model (positive favours Bianchi model).}
\label{tbl:bianchi_evidences}
\centering
\begin{tabular}{lcccc}
\noalign{\doubleline}
{\hfil Model}           & \smica & \sevem & \nilc & \commander \\
\noalign{\vskip 3pt\hrule\vskip 5pt}
Flat-decoupled-Bianchi (left-handed) & $\phantom{-}2.5 \pm 0.1$ & $\phantom{-}3.1 \pm 0.1$ & $\phantom{-}2.3 \pm 0.2$ & $\phantom{-}3.2 \pm 0.2$ \\
Flat-decoupled-Bianchi (right-handed)& $\phantom{-}0.5 \pm 0.1$ & $\phantom{-}0.5 \pm 0.1$ & $\phantom{-}0.5 \pm 0.1$ & $\phantom{-}0.3 \pm 0.1$ \\
Open-coupled-Bianchi (left-handed)   & $\phantom{-}0.2 \pm 0.1$ & $\phantom{-}0.0 \pm 0.1$ & $\phantom{-}0.0 \pm 0.1$ & $\phantom{-}0.0 \pm 0.1$ \\
Open-coupled-Bianchi (right-handed)  & $-0.3 \pm 0.1$           & $-0.5 \pm 0.1$           & $-0.2 \pm 0.1$ & $-0.4 \pm 0.1$ \\
\noalign{\vskip 3pt\hrule\vskip 3pt}
\end{tabular}
\end{table*}

\begin{figure*}
\centering
\begin{subfigure}[b]{170mm}
\includegraphics[viewport=0 0 1 1,clip=,width=0.23\textwidth]{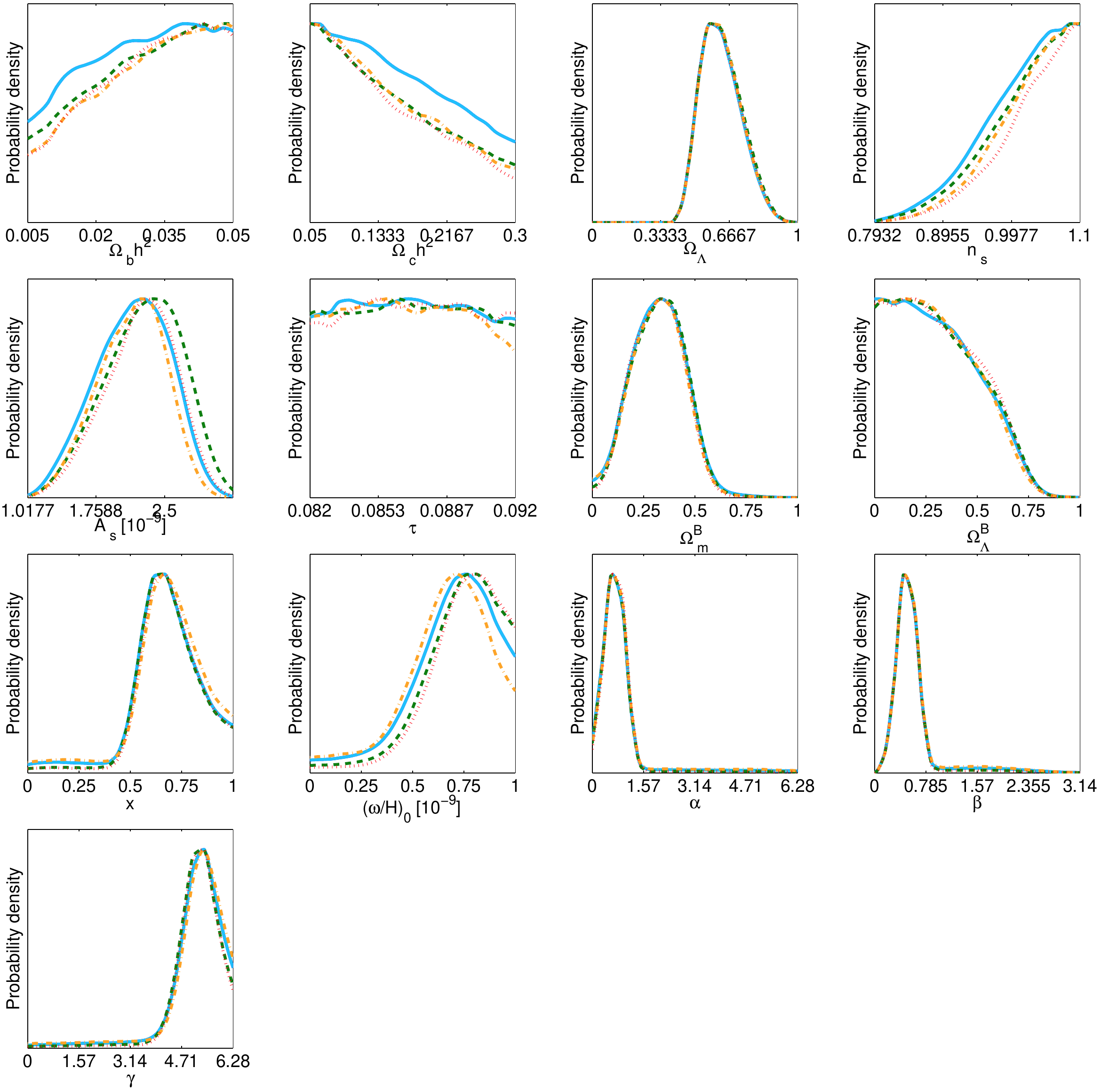}
\includegraphics[viewport=565 550 850 830,clip=,width=0.23\textwidth]{figures/bianchi/dx11_v2/posterior_flat_bianchi_decoupled_lmax032_revised}
\includegraphics[viewport=855 550 1140 830,clip=,width=0.23\textwidth]{figures/bianchi/dx11_v2/posterior_flat_bianchi_decoupled_lmax032_revised}
\includegraphics[viewport=0 270 285 550,clip=,width=0.23\textwidth]{figures/bianchi/dx11_v2/posterior_flat_bianchi_decoupled_lmax032_revised}\\
\includegraphics[viewport=280 270 565 550,clip=,width=0.23\textwidth]{figures/bianchi/dx11_v2/posterior_flat_bianchi_decoupled_lmax032_revised}
\includegraphics[viewport=565 270 850 550,clip=,width=0.23\textwidth]{figures/bianchi/dx11_v2/posterior_flat_bianchi_decoupled_lmax032_revised}
\includegraphics[viewport=855 270 1140 550,clip=,width=0.23\textwidth]{figures/bianchi/dx11_v2/posterior_flat_bianchi_decoupled_lmax032_revised}
\includegraphics[viewport=0 -11 285 269,clip=,width=0.23\textwidth]{figures/bianchi/dx11_v2/posterior_flat_bianchi_decoupled_lmax032_revised}
\caption{Flat-decoupled-Bianchi model.}
\label{fig:posteriors_planck_decoupled}
\end{subfigure} \\*[3mm]
\begin{subfigure}[b]{170mm}
\includegraphics[viewport=850 833 1135 1113,clip=,width=0.23\textwidth]{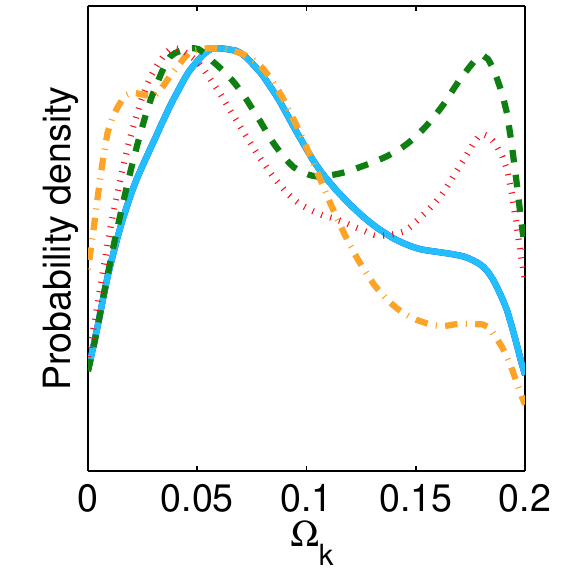}
\includegraphics[viewport=855 550 1140 830,clip=,width=0.23\textwidth]{figures/bianchi/dx11_v2/posterior_open_bianchi_coupled_lmax032_revised}
\includegraphics[viewport=0 270 285 550,clip=,width=0.23\textwidth]{figures/bianchi/dx11_v2/posterior_open_bianchi_coupled_lmax032_revised}
\includegraphics[viewport=280 270 565 550,clip=,width=0.23\textwidth]{figures/bianchi/dx11_v2/posterior_open_bianchi_coupled_lmax032_revised}\\
\includegraphics[viewport=565 270 850 550,clip=,width=0.23\textwidth]{figures/bianchi/dx11_v2/posterior_open_bianchi_coupled_lmax032_revised}
\includegraphics[viewport=855 270 1140 550,clip=,width=0.23\textwidth]{figures/bianchi/dx11_v2/posterior_open_bianchi_coupled_lmax032_revised}
\includegraphics[viewport=0 -11 285 269,clip=,width=0.23\textwidth]{figures/bianchi/dx11_v2/posterior_open_bianchi_coupled_lmax032_revised}
\includegraphics[viewport=280 -11 565 269,clip=,width=0.23\textwidth]{figures/bianchi/dx11_v2/posterior_open_bianchi_coupled_lmax032_revised}
\caption{Open-coupled-Bianchi model.}
\label{fig:posteriors_planck_coupled}
\end{subfigure}
\caption{\rev{Marginalized} posterior distributions of Bianchi parameters recovered from
  \Planck\ \smica\ (solid blue curves), \sevem\ (dashed green curves),
  \nilc\ (dot-dashed yellow curves), and \commander\ (dotted red
  curves) component-separated data for left-handed models.  \Planck\
  data provide evidence in support of a Bianchi component in the
  phenomenological flat-decoupled-Bianchi model (panel a) but not in
  the physical open-coupled-Bianchi model (panel b).  Significant
  differences exist between the posterior distributions shown in panel
  b for each component separation method; this model is not favoured
  by data and parameters are in general poorly constrained.}
\label{fig:posteriors_planck}
\end{figure*}

\begin{figure*}\centering
\mbox{
\begin{subfigure}[b]{.45\textwidth}
  \includegraphics[width=\textwidth]{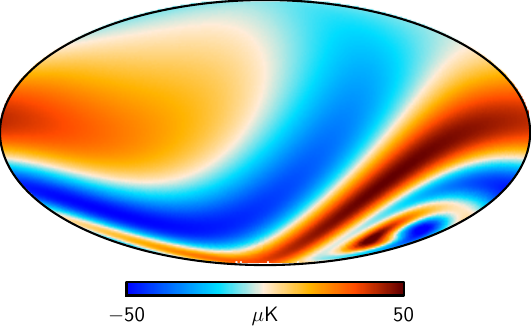}
    \caption{\smica}
\end{subfigure}
\begin{subfigure}[b]{.45\textwidth}
  \includegraphics[width=\textwidth]{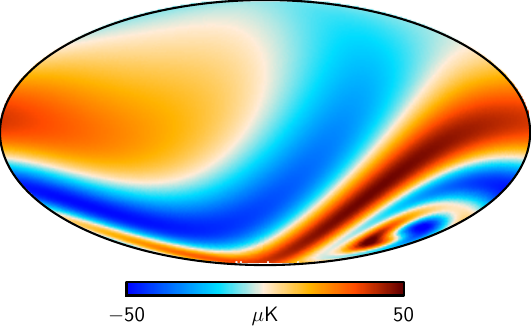}
    \caption{\sevem}
\end{subfigure}
}
\mbox{
\begin{subfigure}[b]{.45\textwidth}
  \includegraphics[width=\textwidth]{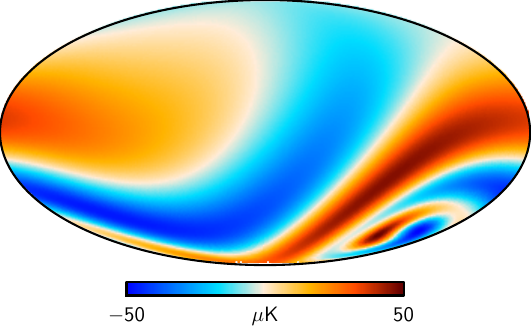}
    \caption{\nilc}
\end{subfigure}
\begin{subfigure}[b]{.45\textwidth}
  \includegraphics[width=\textwidth]{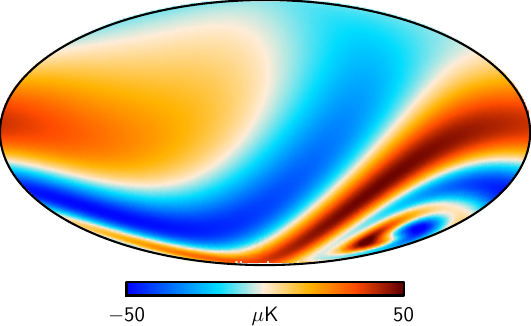}
    \caption{\commander}
\end{subfigure}
}
\caption{Best-fit temperature maps for the left-handed flat-decoupled-Bianchi
  model.}
\label{fig:bianchi_bestfit_decoupled}
\end{figure*}

For the phenomenological flat-decoupled-Bianchi model, evidence in
support of a left-handed Bianchi template is again found
(\tbl{\ref{tbl:bianchi_evidences}}). The Bayes factors
providing evidence for this model range between the values $2.3 \pm
0.2$ and \mbox{$3.2 \pm 0.2$}, corresponding to odds ratios of
approximately 1:10 and 1:25, respectively (which on the Jeffreys scale
are categorized as significant and strong, respectively;
\citealt{jeffreys:1961}). Recovered posterior distributions of the
Bianchi parameters of this model for each component separation
technique are shown in \fig{\ref{fig:posteriors_planck_decoupled}}.
The posterior distributions are consistent across component
separation techniques, are similar to those recovered in
\citet{planck2013-p19}, and are reasonably well constrained (with the
exception of the known $\Omega_{\rm m}^{\rm
  B}$--$\Omega_{\Lambda}^{\rm B}$ Bianchi parameter degeneracy
\citep{jaffe:2006b,bridges:2006b}). Recall that the Bianchi
densities ($\Omega_{\rm m}^{\rm B}$ and $\Omega_{\Lambda}^{\rm
  B}$) are decoupled from the standard cosmology in the
flat-decoupled-Bianchi model considered here and, as found previously,
are inconsistent with standard estimates of the densities.
The maximum-a-posteriori (MAP) and mean-posterior Bianchi parameter
estimates for this model are given in
\tbl{\ref{tbl:bianchi_bestfit_parameters_decoupled}}, while the
corresponding MAP best-fit Bianchi temperature maps are shown in
\fig{\ref{fig:bianchi_bestfit_decoupled}}. 
\rev{Note that the maximum of a marginalized one-dimensional posterior (e.g.,  \fig{\ref{fig:posteriors_planck}}) will not in general coincide with the global MAP estimate for the full set of parameters.}
The best-fit maps for the
left-handed flat-coupled-Bianchi model are consistent across component
separation techniques and similar to the best-fit maps found in
previous \Planck\ \citep{planck2013-p19} and {WMAP}
(see, e.g., \citealt{jaffe:2005} and \citealt{mcewen:bianchi}) temperature data.

\begin{figure*}\centering
\mbox{
\begin{subfigure}[b]{.45\textwidth}
  \includegraphics[width=\textwidth]{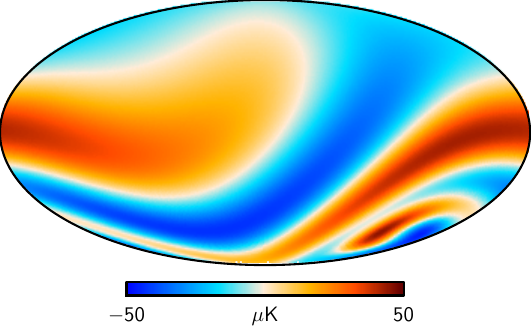}
    \caption{\smica}
\end{subfigure}
\begin{subfigure}[b]{.45\textwidth}
  \includegraphics[width=\textwidth]{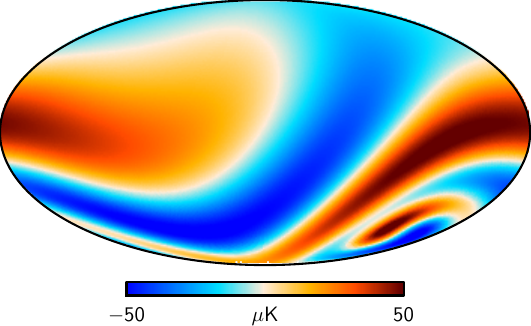}
    \caption{\sevem}
\end{subfigure}
}
\mbox{
\begin{subfigure}[b]{.45\textwidth}
  \includegraphics[width=\textwidth]{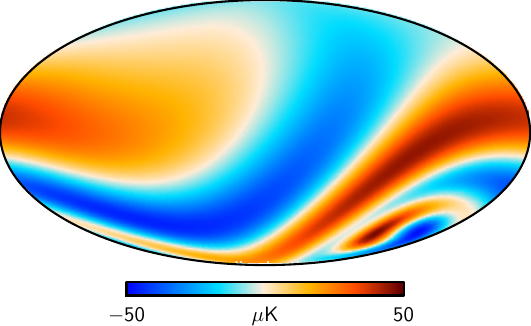}
    \caption{\nilc}
\end{subfigure}
\begin{subfigure}[b]{.45\textwidth}
  \includegraphics[width=\linewidth]{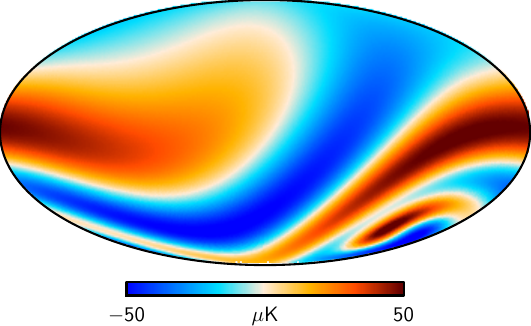}
    \caption{\commander}
\end{subfigure}
}
\caption{Best-fit temperature maps for the left-handed open-coupled-Bianchi model.}
\label{fig:bianchi_bestfit_coupled}
\end{figure*}

For the physical open-coupled-Bianchi model where the \bianchiviih\
model is coupled to the standard cosmology, there is again no evidence
in support of a Bianchi contribution
(\tbl{\ref{tbl:bianchi_evidences}}). Recovered posterior distributions
of the parameters of this model for each component separation
technique are shown in \fig{\ref{fig:posteriors_planck_coupled}}. The
posterior distributions show some similarity across component
separation techniques and with the distributions recovered in
\citet{planck2013-p19}.  However, significant differences exist since
the parameters are in general poorly constrained and the model is not
favoured by the Bayesian evidence.
The MAP and mean-posterior Bianchi parameter estimates for this model
are given in \tbl{\ref{tbl:bianchi_bestfit_parameters_coupled}}, while
the corresponding MAP best-fit Bianchi temperature maps are shown in
\fig{\ref{fig:bianchi_bestfit_coupled}}. While the posterior
distributions show differences between component separation
techniques, the estimated parameters are consistent. Note that,
although the mean parameter estimates have not changed markedly from
\citet{planck2013-p19}, the MAP estimates have indeed changed.
Intriguingly, for each component separation technique, the MAP
best-fit maps for the open-coupled-Bianchi model
(\fig{\ref{fig:bianchi_bestfit_coupled}}), for which there is no
evidence in support of a Bianchi contribution, show a similar but not
identical morphology to the MAP best-fit maps for the
flat-decoupled-Bianchi model
(\fig{\ref{fig:bianchi_bestfit_decoupled}}), for which the Bayesian
evidence supports the inclusion of a Bianchi component. This was not
the case in the previous \Planck\ analysis
\citep[][\fig{22}]{planck2013-p19}.  A parameter combination is found
for the physical open-coupled-Bianchi model that is broadly consistent
with a standard open cosmology and that produces a Bianchi temperature
map similar to the one found in the unphysical flat-decoupled-Bianchi
model.  This parameter combination lies on the known $\Omega_{\rm
  m}^{\rm B}$--$\Omega_{\Lambda}^{\rm B}$ Bianchi parameter degeneracy
\citep{jaffe:2006b,bridges:2006b} and lies close to but not directly on
the well-known CMB geometric degeneracy (determined by an independent
CMB analysis), since both the cosmological and Bianchi components are
fitted simultaneously.  It is important to stress that \Planck\
temperature data do not favour \rev{the physical open-coupled-Bianchi model, but neither is it possible to rule out this model using temperature data alone (\Planck\ polarization data are considered subsequently).}
An overall constraint on the vorticity of \bianchiviih\ models, \rev{from \Planck\ temperature data alone}, of
\mbox{$(\omega/H)_0 < 7.6 \times 10^{-10}$} (95\,\% confidence level) is
obtained from the analysis of the physical open-coupled-Bianchi model,
which is consistent across all component separation techniques, as
illustrated in \tbl{\ref{tbl:bianchi_vorticity_bounds}}.

\begin{figure*}\centering
\mbox{
\begin{subfigure}[b]{.48\textwidth}
  \includegraphics[width=.48\textwidth]{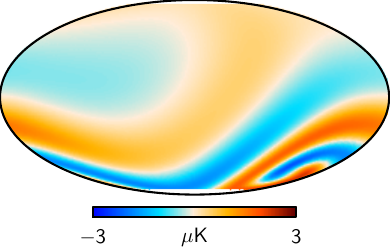}
  \includegraphics[width=.48\textwidth]{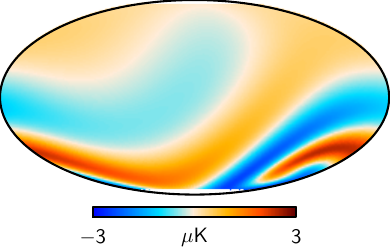}
    \caption{\smica}
\end{subfigure}
\begin{subfigure}[b]{.48\textwidth}
  \includegraphics[width=.48\textwidth]{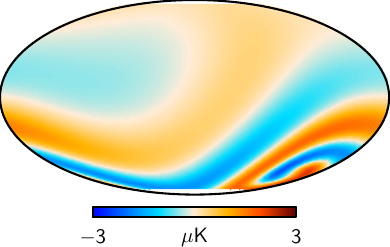}
  \includegraphics[width=.48\textwidth]{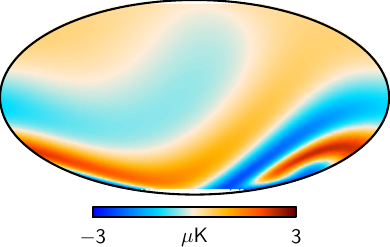}
    \caption{\sevem}
\end{subfigure}
}
\mbox{
\begin{subfigure}[b]{.48\textwidth}
  \includegraphics[width=.48\textwidth]{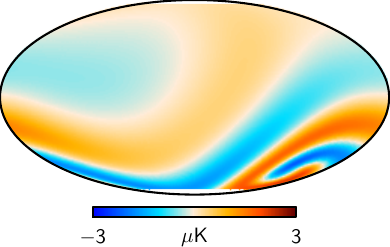}
  \includegraphics[width=.48\textwidth]{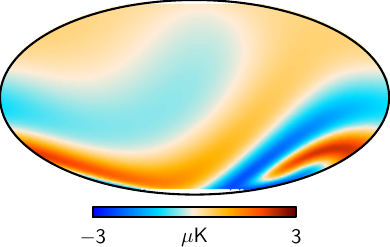}
    \caption{\nilc}
\end{subfigure}
\begin{subfigure}[b]{.48\textwidth}
  \includegraphics[width=.48\textwidth]{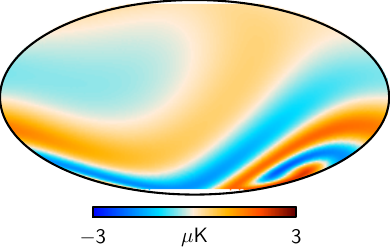}
  \includegraphics[width=.48\textwidth]{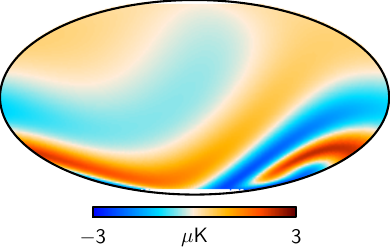}
    \caption{\commander}
\end{subfigure}
}
\caption{Polarization maps for the best-fit left-handed
  flat-decoupled-Bianchi model fitted to temperature data.  In each
  panel $E$-~(left) and $B$-mode~(right) maps are shown. These
  polarization maps are simulated using the approach of
  \citet{pontzen:2007} and \citet{pontzen:2009}, and provided by \citet{PontzenPrivate}.
}
\label{fig:bianchi_polarization_decoupled}
\end{figure*}

\begin{figure*}\centering
\mbox{
\begin{subfigure}[b]{.48\textwidth}
  \includegraphics[width=.48\textwidth]{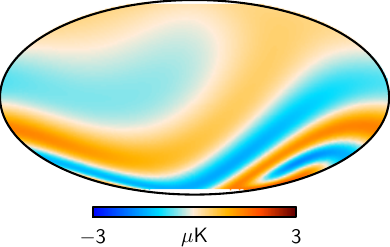}
  \includegraphics[width=.48\textwidth]{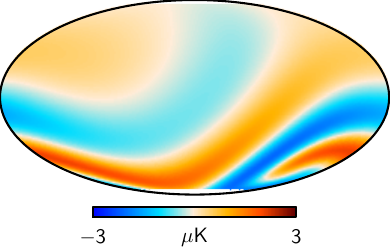}
    \caption{\smica}
\end{subfigure}
\begin{subfigure}[b]{.48\textwidth}
  \includegraphics[width=.48\textwidth]{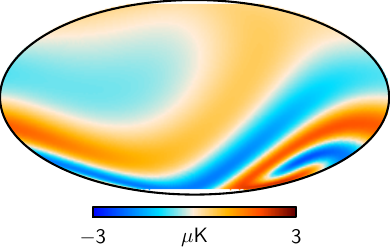}
  \includegraphics[width=.48\textwidth]{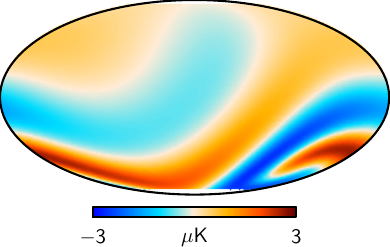}
    \caption{\sevem}
\end{subfigure}
}
\mbox{
\begin{subfigure}[b]{.48\textwidth}
  \includegraphics[width=.48\textwidth]{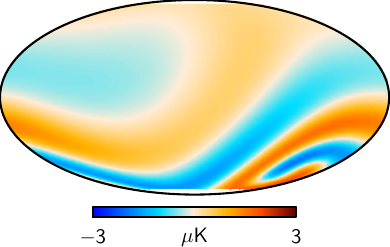}
  \includegraphics[width=.48\textwidth]{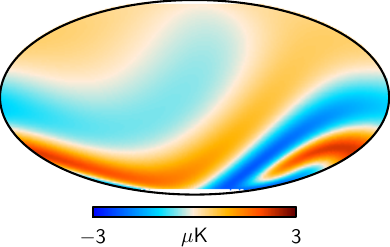}
    \caption{\nilc}
\end{subfigure}
\begin{subfigure}[b]{.48\textwidth}
  \includegraphics[width=.48\textwidth]{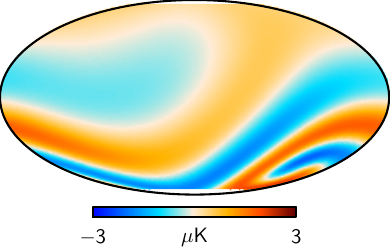}
  \includegraphics[width=.48\textwidth]{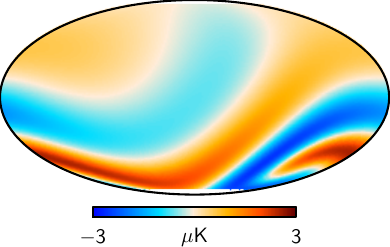}
    \caption{\commander}
\end{subfigure}
}
\caption{Polarization maps for the best-fit left-handed
  open-coupled-Bianchi model fitted to temperature data.  In each
  panel $E$-~(left) and $B$-mode~(right) maps are shown. These
  polarization maps are simulated using the approach of
  \citet{pontzen:2007} and \citet{pontzen:2009}, and provided by \citet{PontzenPrivate}.
}
\label{fig:bianchi_polarization_coupled}
\end{figure*}

To further constrain \bianchiviih\ models using \Planck\ polarization
data we simulate Bianchi polarization maps, computed using the
approach of \citet{pontzen:2007} and \citet{pontzen:2009}, and provided
by \citet{PontzenPrivate}
for the best-fit Bianchi
parameters determined from the temperature analysis.  $E$- and
$B$-mode Bianchi maps for the best-fit flat-decoupled-Bianchi model
and the open-coupled-Bianchi model are displayed in
\fig{\ref{fig:bianchi_polarization_decoupled}} and
\fig{\ref{fig:bianchi_polarization_coupled}}, respectively.  As
described in Sect.~\ref{subs:bianchi}, we estimate the
maximum-likelihood amplitude of these Bianchi polarization templates
in \Planck\ polarization data, performing the analysis in the pixel
space defined by the $Q$ and $U$ Stokes components, where noise and
partial sky coverage can be handled effectively.  The
maximum-likelihood amplitudes estimated for each component separation
technique are given in
\tbl{\ref{tbl:bianchi_polarization_amplitudes}}.  The estimated
amplitudes are close to zero and consistent across component
separation techniques. The difference between the estimated
amplitudes and zero is more likely due to small residual foreground
contamination than a Bianchi component.  Indeed, the amplitude
estimates are at least 24 standard deviations from unity, the expected
value for the best-fit Bianchi models determined from the temperature
analysis.  Both the best-fit flat-decoupled-Bianchi model and the
open-decoupled-Bianchi model are thus strongly disfavoured by the
\Planck\ polarization data.  This is not surprising since these models
produce relatively large $E$- \emph{and} $B$-mode contributions (see
\fig{\ref{fig:bianchi_polarization_decoupled}} and
\fig{\ref{fig:bianchi_polarization_coupled}}), as highlighted already
by \citet{pontzen:2007}.  However, the full freedom of Bianchi models
remains to be explored using temperature and polarization data
simultaneously, for example through a complete Bayesian analysis,
which is left to future work.

\begin{table*}
\caption{Parameters recovered for the left-handed flat-decoupled-Bianchi
  model. \Planck\ data favour the inclusion of a Bianchi component in
  this phenomenological model.}
\label{tbl:bianchi_bestfit_parameters_decoupled}
\centering
\begin{tabular}{lccccccccc}
\noalign{\doubleline}
\multicolumn{1}{c}{Bianchi}     & \multicolumn{2}{c}{\smica} & \multicolumn{2}{c}{\sevem} & \multicolumn{2}{c}{\nilc} & \multicolumn{2}{c}{\commander} \\
\multicolumn{1}{c}{parameter}   & MAP & Mean & MAP & Mean & MAP & Mean & MAP & Mean \\
\noalign{\vskip 3pt\hrule\vskip 5pt}
\hfil$\Omega_{\rm m}^{\rm B}$ & $0.34$ & $0.32 \pm 0.12$ & $0.34$ & $0.33 \pm 0.12$
& $0.42$ & $0.32 \pm 0.12$
 & $0.23$ & $0.32 \pm 0.12$ \\
\hfil$\Omega_{\Lambda}^{\rm B}$  & $0.30$ & $0.31 \pm 0.20$ & $0.26$ & $0.31 \pm 0.20$
& $0.12$ & $0.30 \pm 0.19$
 & $0.48$ & $0.31 \pm 0.20$ \\
\hfil$\bx$ & $0.66$ & $0.67 \pm 0.17$ & $0.63$ & $0.68 \pm 0.15$
& $0.63$ & $0.68 \pm 0.18$
& $0.74$ & $0.69 \pm 0.15$ \\
\hfil$(\omega/H)_0 \times 10^{10}$  & $ 8.6$ & $6.9 \pm  2.0$  & $ 9.2$ & $7.3 \pm  1.8$
 & $ 8.2$ & $6.6 \pm  2.0$ & $ 8.2$ & $7.4 \pm  1.7$ \\
\hfil$\eula$  & $ 39\pdeg 4$ & $ 52\pdeg 7 \pm  49\pdeg 7$  & $ 39\pdeg 3$ & $ 48\pdeg 0 \pm  39\pdeg 5$ & $ 41\pdeg 1$ & $ 57\pdeg 1 \pm  56\pdeg 0$ & $ 41\pdeg 9$ & $ 47\pdeg 8 \pm  36\pdeg 1$ \\
\hfil$\eulb$  & $ 27\pdeg 8$ & $ 34\pdeg 4 \pm  21\pdeg 1$  & $ 28\pdeg 2$ & $ 32\pdeg 2 \pm  16\pdeg 8$ & $ 28\pdeg 9$ & $ 35\pdeg 8 \pm  22\pdeg 8$ & $ 27\pdeg 6$ & $ 31\pdeg 5 \pm  15\pdeg 4$\\
\hfil$\eulc$  & $302\pdeg 1$ & $291\pdeg 5 \pm  53\pdeg 4\phantom{0}$  & $309\pdeg 2$ & $293\pdeg 3 \pm  44\pdeg 7\phantom{0}$ & $297\pdeg 6$ & $291\pdeg 2 \pm  59\pdeg 0\phantom{0}$ & $303\pdeg 5$ & $295\pdeg 3 \pm  41\pdeg 3\phantom{0}$ \\
\noalign{\vskip 3pt\hrule\vskip 3pt}
\end{tabular}
\end{table*}

\begin{table*}
\caption{Parameters recovered for the left-handed open-coupled-Bianchi model. \Planck\ data do not favour the inclusion of a Bianchi component in this model and some parameters are not well constrained.}
\label{tbl:bianchi_bestfit_parameters_coupled}
\centering
\begin{tabular}{lccccccccc}
\noalign{\doubleline}
\multicolumn{1}{c}{Bianchi}     & \multicolumn{2}{c}{\smica} & \multicolumn{2}{c}{\sevem} & \multicolumn{2}{c}{\nilc} & \multicolumn{2}{c}{\commander} \\
\multicolumn{1}{c}{parameter}   & MAP & Mean & MAP & Mean & MAP & Mean & MAP & Mean \\
\noalign{\vskip 3pt\hrule\vskip 5pt}
\hfil$\Omega_{k}$ & $0.20$ & $0.09 \pm 0.05$ & $0.18$ & $0.10 \pm 0.06$
& $0.16$ & $0.08 \pm 0.05$ & $0.19$ & $0.10 \pm 0.06$ \\
\hfil$\Omega_{\rm m}^{\rm B}$ & $0.23$ & $0.31 \pm 0.07$ & $0.20$ & $0.30 \pm 0.07$
& $0.16$ & $0.32 \pm 0.07$ & $0.24$ & $0.30 \pm 0.08$ \\
\hfil$\Omega_{\Lambda}^{\rm B}$ & $0.57$ & $0.60 \pm 0.07$  & $0.62$ & $0.60 \pm 0.07$
& $0.67$ & $0.60 \pm 0.07$ & $0.57$ & $0.61 \pm 0.07$ \\
\hfil$\bx$ & $0.74$ & $0.45 \pm 0.28$ & $0.87$ & $0.51 \pm 0.28$
& $0.93$ & $0.42 \pm 0.29$ & $0.78$ & $0.49 \pm 0.28$ \\
\hfil$(\omega/H)_0 \times 10^{10}$ & $ 6.2$ & $3.8 \pm  2.4$ & $ 6.5$ & $4.1 \pm  2.3$
& $ 5.4$ & $3.3 \pm  2.3$ & $ 6.8$ & $3.9 \pm  2.3$ \\
\hfil$\eula$ & $ 39\pdeg 0$ & $136\pdeg 8 \pm 100\pdeg 6$ & $ 41\pdeg 0$ & $116\pdeg 1 \pm  96\pdeg 6\phantom{0}$ & $ 43\pdeg 4$ & $161\pdeg 2 \pm 101\pdeg 7$ & $ 40\pdeg 1$ & $121\pdeg 5 \pm  96\pdeg 8\phantom{0}$ \\
\hfil$\eulb$ & $ 27\pdeg 6$ & $ 72\pdeg 3 \pm  38\pdeg 8$ & $ 29\pdeg 4$ & $ 63\pdeg 2 \pm  39\pdeg 1$ & $ 28\pdeg 4$ & $ 83\pdeg 3 \pm  37\pdeg 4$ & $ 28\pdeg 4$ & $ 66\pdeg 6 \pm  38\pdeg 8$ \\
\hfil$\eulc$ & $264\pdeg 7$ & $194\pdeg 1 \pm  87\pdeg 3\phantom{0}$ & $272\pdeg 0$ & $210\pdeg 2 \pm  80\pdeg 1\phantom{0}$ & $289\pdeg 7$ & $177\pdeg 7 \pm  90\phantom{0}\pdeg 9$ & $262\pdeg 6$ & $201\pdeg 5 \pm  82\pdeg 2\phantom{0}$ \\
\noalign{\vskip 3pt\hrule\vskip 3pt}
\end{tabular}
\end{table*}

\begin{table*}
\caption{Upper bounds on vorticity $(\omega/H)_0$ at 95\,\% confidence level.}
\label{tbl:bianchi_vorticity_bounds}
\centering
\begin{tabular}{lcccc} 
\noalign{\doubleline}
{\hfil Model}           & \smica & \sevem & \nilc & \commander \\
\noalign{\vskip 3pt\hrule\vskip 5pt}
Open-coupled-Bianchi (left-handed)   & $7.6 \times 10^{-10}$ &  $7.6 \times 10^{-10}$ &  $7.6 \times 10^{-10}$ &  $7.6 \times 10^{-10}$ \\
Open-coupled-Bianchi (right-handed)  & $7.6 \times 10^{-10}$ &  $7.6 \times 10^{-10}$ &  $7.6 \times 10^{-10}$ &  $7.1 \times 10^{-10}$ \\
\noalign{\vskip 3pt\hrule\vskip 3pt}
\end{tabular}
\end{table*}

\begin{table*}
  \caption{Maximum-likelihood amplitude estimates $\lambda^{\rm ML}$ and
    1\,$\sigma$ errors computed using polarization data.}
\label{tbl:bianchi_polarization_amplitudes}
\centering
\begin{tabular}{lcccc} 
\noalign{\doubleline}
{\hfil Model}           & \smica & \sevem & \nilc & \commander \\
\noalign{\vskip 3pt\hrule\vskip 5pt}
Flat-decoupled-Bianchi (left-handed) & $-0.10 \pm 0.04$ & $-0.10 \pm 0.04$ & $-0.12 \pm 0.04$ & $-0.11 \pm 0.04$ \\
Open-coupled-Bianchi (left-handed)   & $-0.11 \pm 0.04$ & $-0.09 \pm 0.04$ & $-0.11 \pm 0.04$ & $-0.09 \pm 0.04$ \\
\noalign{\vskip 3pt\hrule\vskip 3pt}
\end{tabular}
\end{table*}

\section{Discussion} 
\label{sec:discussion}

We have used \Planck\ intensity and polarization data to evaluate
specific departures from the large-scale isotropy of the
Universe. Using both frequentist and Bayesian methods applied for the first time to polarization data,
we find no
evidence for a multi-connected topology with a scale less than roughly
the distance to the last-scattering surface. Specifically, a
frequentist search for antipodal matched circles on
$N_\mathrm{side}=512$ maps  finds a lower bound on the size of the
fundamental domain of $0.97\,\chi_\mathrm{rec}$ from polarization data alone. Using Bayesian methods applied to low-resolution ($N_\mathrm{side}=16$) maps of both temperature and polarization, we also find a  lower limit of $0.97\,\chi_\mathrm{rec}$ for the T3 cubic torus (for the T1 slab, the limit is $0.56\,\chi_\mathrm{rec}$). These results are both consistent and complementary, giving coincidentally identical limits but with very different statistical foundations and data selections. The addition of polarization data at current levels of accuracy does not significantly improve the limits from intensity alone, but we have found that the polarization sensitivity of the full set of \Planck\ detectors should give quantitative improvements in the limits, decreasing the likelihood of fundamental domains with scales smaller than the distance to the last-scattering surface by many orders of magnitude.

We also find no evidence for a \bianchiviih\ model which departs from global isotropy via the presence of both rotation and shear. Although the large-scale temperature pattern measured by \Planck\ has some similar features to that induced by focusing in a \bianchiviih\ universe, it requires unphysical parameters. Fixed to those parameters, we have further shown that the polarization pattern induced by such models is strongly disfavoured by the \Planck\ data.

\rev{The results outlined here show no evidence for departures from isotropic and simply-connected models. Improved computational techniques, along with future polarization data from \Planck\ (and beyond), will allow yet stronger checks of even wider classes of models. For the multi-connected case, we can expand to models without antipodal matched circles and with the scale of the fundamental domain closer to (and even slightly beyond) the last-scattering surface. For the likelihood method, this will require computation of the correlation matrix with higher-wavenumber modes to capture more of the available information.}

\rev{For anisotropic models, we can explicitly perform parameter estimation and model selection using polarization data, beyond the simple template-fitting performed here.}

\rev{Although the evidence thus far corroborates the conventional wisdom that we live in the simplest FLRW Universe, this is likely to be only an approximation vastly beyond the Hubble scale. Detection of a multiply-connected topology or anisotropic geometry is one of the few ways to probe the global structure of spacetime. We have shown that \Planck\ data give the best handle to date on these possibilities. }



\begin{acknowledgements}

The Planck Collaboration acknowledges the support of: ESA; CNES and CNRS/INSU-IN2P3-INP (France); ASI, CNR, and INAF (Italy); NASA and DoE (USA); STFC and UKSA (UK); CSIC, MINECO, JA, and RES (Spain); Tekes, AoF, and CSC (Finland); DLR and MPG (Germany); CSA (Canada); DTU Space (Denmark); SER/SSO (Switzerland); RCN (Norway); SFI (Ireland); FCT/MCTES (Portugal); ERC and PRACE (EU). A description of the Planck Collaboration and a list of its members, indicating which technical or scientific activities they have been involved in, can be found at 
\href{http://www.cosmos.esa.int/web/planck/planck-collaboration}{\texttt{http://www.cosmos.esa.int/web/planck/planck-collaboration}}.
The authors thank Andrew Pontzen for computing \bianchiviih\ polarization
templates for the best-fit models resulting from the analysis of
temperature data.  We acknowledge the UCL Legion High Performance Computing Facility
(Legion@UCL) and associated support services in the completion of
this work. Parts of the computations were
performed on the Andromeda and Perseus clusters of the University of Geneva,
as well as the Carver IBM iDataPlex, the Hopper Cray XE6, and the Edison Cray XC30 at NERSC, and on
the GPC supercomputer at the SciNet HPC Consortium. SciNet is funded by: the Canada Foundation for
Innovation under the auspices of Compute Canada; the Government of Ontario; Ontario Research Fund -
Research Excellence; and the University of Toronto.

\end{acknowledgements}

\bibliography{topo,Planck_bib}

\raggedright 
\end{document}

%% file: Planck.tex
\def\setsymbol#1#2{\expandafter\def\csname #1\endcsname{#2}}
\def\getsymbol#1{\csname #1\endcsname}

\def\Planck{\textit{Planck}}

\def\alltwentyfifteenresultspapers{\nocite{planck2014-a01, planck2014-a03, planck2014-a04, planck2014-a05, planck2014-a06, planck2014-a07, planck2014-a08, planck2014-a09, planck2014-a11, planck2014-a12, planck2014-a13, planck2014-a14, planck2014-a15, planck2014-a16, planck2014-a17, planck2014-a18, planck2014-a19, planck2014-a20, planck2014-a22, planck2014-a24, planck2014-a26, planck2014-a28, planck2014-a29, planck2014-a30, planck2014-a31, planck2014-a35, planck2014-a36, planck2014-a37, planck2014-ES}}

\newbox\tablebox    \newdimen\tablewidth
\def\leaderfil{\leaders\hbox to 5pt{\hss.\hss}\hfil}
\def\tablenote#1 #2\par{\begingroup \parindent=0.8em
    \abovedisplayshortskip=0pt\belowdisplayshortskip=0pt
    \noindent
    $$\hss\vbox{\hsize\tablewidth \hangindent=\parindent \hangafter=1 \noindent
    \hbox to \parindent{$^#1$\hss}\strut#2\strut\par}\hss$$
    \endgroup}
\def\doubleline{\vskip 3pt\hrule \vskip 1.5pt \hrule \vskip 5pt}

\def\L2{\ifmmode L_2\else $L_2$\fi}

\def\DeltaT{\ifmmode \Delta T\else $\Delta T$\fi}
\def\deltat{\ifmmode \Delta t\else $\Delta t$\fi}
\def\fknee{\ifmmode f_{\rm knee}\else $f_{\rm knee}$\fi}
\def\Fmax{\ifmmode F_{\rm max}\else $F_{\rm max}$\fi}
\def\solar{\ifmmode{\rm M}_{\mathord\odot}\else${\rm M}_{\mathord\odot}$\fi}
\def\Msolar{\ifmmode{\rm M}_{\mathord\odot}\else${\rm M}_{\mathord\odot}$\fi}
\def\Lsolar{\ifmmode{\rm L}_{\mathord\odot}\else${\rm L}_{\mathord\odot}$\fi}
\def\inv{\ifmmode^{-1}\else$^{-1}$\fi}
\def\mo{\ifmmode^{-1}\else$^{-1}$\fi}
\def\sup#1{\ifmmode ^{\rm #1}\else $^{\rm #1}$\fi}
\def\expo#1{\ifmmode \times 10^{#1}\else $\times 10^{#1}$\fi}
\def\,{\thinspace}
\def\lsim{\mathrel{\raise .4ex\hbox{\rlap{$<$}\lower 1.2ex\hbox{$\sim$}}}}
\def\gsim{\mathrel{\raise .4ex\hbox{\rlap{$>$}\lower 1.2ex\hbox{$\sim$}}}}

\def\simprop{\mathrel{\raise .4ex\hbox{\rlap{$\propto$}\lower 1.2ex\hbox{$\sim$}}}}
\def\deg{\ifmmode^\circ\else$^\circ$\fi}
\def\pdeg{\ifmmode $\setbox0=\hbox{$^{\circ}$}\rlap{\hskip.11\wd0 .}$^{\circ}
          \else \setbox0=\hbox{$^{\circ}$}\rlap{\hskip.11\wd0 .}$^{\circ}$\fi}
\def\arcs{\ifmmode {^{\scriptstyle\prime\prime}}
          \else $^{\scriptstyle\prime\prime}$\fi}
\def\arcm{\ifmmode {^{\scriptstyle\prime}}
          \else $^{\scriptstyle\prime}$\fi}
\newdimen\sa  \newdimen\sb
\def\parcs{\sa=.07em \sb=.03em
     \ifmmode \hbox{\rlap{.}}^{\scriptstyle\prime\kern -\sb\prime}\hbox{\kern -\sa}
     \else \rlap{.}$^{\scriptstyle\prime\kern -\sb\prime}$\kern -\sa\fi}
\def\parcm{\sa=.08em \sb=.03em
     \ifmmode \hbox{\rlap{.}\kern\sa}^{\scriptstyle\prime}\hbox{\kern-\sb}
     \else \rlap{.}\kern\sa$^{\scriptstyle\prime}$\kern-\sb\fi}
\def\ra[#1 #2 #3.#4]{#1\sup{h}#2\sup{m}#3\sup{s}\llap.#4}
\def\dec[#1 #2 #3.#4]{#1\deg#2\arcm#3\arcs\llap.#4}
\def\deco[#1 #2 #3]{#1\deg#2\arcm#3\arcs}
\def\rra[#1 #2]{#1\sup{h}#2\sup{m}}

\def\dots{\relax\ifmmode \ldots\else $\ldots$\fi}
\def\WHzsr{\ifmmode $W\,Hz\mo\,sr\mo$\else W\,Hz\mo\,sr\mo\fi}
\def\mHz{\ifmmode $\,mHz$\else \,mHz\fi}
\def\GHz{\ifmmode $\,GHz$\else \,GHz\fi}
\def\mKs{\ifmmode $\,mK\,s$^{1/2}\else \,mK\,s$^{1/2}$\fi}
\def\muKs{\ifmmode \,\mu$K\,s$^{1/2}\else \,$\mu$K\,s$^{1/2}$\fi}
\def\muKRJs{\ifmmode \,\mu$K$_{\rm RJ}$\,s$^{1/2}\else \,$\mu$K$_{\rm RJ}$\,s$^{1/2}$\fi}
\def\muKHz{\ifmmode \,\mu$K\,Hz$^{-1/2}\else \,$\mu$K\,Hz$^{-1/2}$\fi}
\def\MJysr{\ifmmode \,$MJy\,sr\mo$\else \,MJy\,sr\mo\fi}
\def\MJysrmK{\ifmmode \,$MJy\,sr\mo$\,mK$_{\rm CMB}\mo\else \,MJy\,sr\mo\,mK$_{\rm CMB}\mo$\fi}
\def\microns{\ifmmode \,\mu$m$\else \,$\mu$m\fi}

\def\muK{\ifmmode \,\mu$K$\else \,$\mu$\hbox{K}\fi}
\def\microK{\ifmmode \,\mu$K$\else \,$\mu$\hbox{K}\fi}
\def\muW{\ifmmode \,\mu$W$\else \,$\mu$\hbox{W}\fi}
\def\kms{\ifmmode $\,km\,s$^{-1}\else \,km\,s$^{-1}$\fi}
\def\kmsMpc{\ifmmode $\,\kms\,Mpc\mo$\else \,\kms\,Mpc\mo\fi}
\providecommand{\sorthelp}[1]{}

%% file: macros_data.tex
\newcommand{\mksym}[1]{\ifmmode {\rm #1}\else #1\fi}

\newcommand{\lowEB}{\mksym{{\rm lowP}}}
\newcommand{\lowTP}{\mksym{{\rm lowT{,}P}}}

%% file: A20_Geometry_topology_authors_and_institutes.tex
\author{\small
Planck Collaboration: P.~A.~R.~Ade\inst{95}
\and
N.~Aghanim\inst{62}
\and
M.~Arnaud\inst{78}
\and
M.~Ashdown\inst{74, 6}
\and
J.~Aumont\inst{62}
\and
C.~Baccigalupi\inst{93}
\and
A.~J.~Banday\inst{105, 10}
\and
R.~B.~Barreiro\inst{69}
\and
N.~Bartolo\inst{33, 70}
\and
S.~Basak\inst{93}
% ...
}
\institute{\small
APC, AstroParticule et Cosmologie, Universit\'{e} Paris Diderot, CNRS/IN2P3, CEA/lrfu, Observatoire de Paris, Sorbonne Paris Cit\'{e}, 10, rue Alice Domon et L\'{e}onie Duquet, 75205 Paris Cedex 13, France\goodbreak
\and
Aalto University Mets\"{a}hovi Radio Observatory and Dept of Radio Science and Engineering, P.O. Box 13000, FI-00076 AALTO, Finland\goodbreak
\and
African Institute for Mathematical Sciences, 6-8 Melrose Road, Muizenberg, Cape Town, South Africa\goodbreak
\and
Agenzia Spaziale Italiana Science Data Center, Via del Politecnico snc, 00133, Roma, Italy\goodbreak
\and
Aix Marseille Universit\'{e}, CNRS, LAM (Laboratoire d'Astrophysique de Marseille) UMR 7326, 13388, Marseille, France\goodbreak
\and
Astrophysics Group, Cavendish Laboratory, University of Cambridge, J J Thomson Avenue, Cambridge CB3 0HE, U.K.\goodbreak
\and
Astrophysics \& Cosmology Research Unit, School of Mathematics, Statistics \& Computer Science, University of KwaZulu-Natal, Westville Campus, Private Bag X54001, Durban 4000, South Africa\goodbreak
\and
CGEE, SCS Qd 9, Lote C, Torre C, 4$^{\circ}$ andar, Ed. Parque Cidade Corporate, CEP 70308-200, Bras\'{i}lia, DF, Brazil\goodbreak
\and
CITA, University of Toronto, 60 St. George St., Toronto, ON M5S 3H8, Canada\goodbreak
\and
CNRS, IRAP, 9 Av. colonel Roche, BP 44346, F-31028 Toulouse cedex 4, France\goodbreak
\and
CRANN, Trinity College, Dublin, Ireland\goodbreak
\and
California Institute of Technology, Pasadena, California, U.S.A.\goodbreak
\and
Centre for Theoretical Cosmology, DAMTP, University of Cambridge, Wilberforce Road, Cambridge CB3 0WA, U.K.\goodbreak
\and
Centro de Estudios de F\'{i}sica del Cosmos de Arag\'{o}n (CEFCA), Plaza San Juan, 1, planta 2, E-44001, Teruel, Spain\goodbreak
\and
Computational Cosmology Center, Lawrence Berkeley National Laboratory, Berkeley, California, U.S.A.\goodbreak
\and
Consejo Superior de Investigaciones Cient\'{\i}ficas (CSIC), Madrid, Spain\goodbreak
\and
DSM/Irfu/SPP, CEA-Saclay, F-91191 Gif-sur-Yvette Cedex, France\goodbreak
\and
DTU Space, National Space Institute, Technical University of Denmark, Elektrovej 327, DK-2800 Kgs. Lyngby, Denmark\goodbreak
\and
D\'{e}partement de Physique Th\'{e}orique, Universit\'{e} de Gen\`{e}ve, 24, Quai E. Ansermet,1211 Gen\`{e}ve 4, Switzerland\goodbreak
\and
Departamento de Astrof\'{i}sica, Universidad de La Laguna (ULL), E-38206 La Laguna, Tenerife, Spain\goodbreak
\and
Departamento de F\'{\i}sica, Universidad de Oviedo, Avda. Calvo Sotelo s/n, Oviedo, Spain\goodbreak
\and
Department of Astronomy and Astrophysics, University of Toronto, 50 Saint George Street, Toronto, Ontario, Canada\goodbreak
\and
Department of Astrophysics/IMAPP, Radboud University Nijmegen, P.O. Box 9010, 6500 GL Nijmegen, The Netherlands\goodbreak
\and
Department of Physics \& Astronomy, University of British Columbia, 6224 Agricultural Road, Vancouver, British Columbia, Canada\goodbreak
\and
Department of Physics and Astronomy, Dana and David Dornsife College of Letter, Arts and Sciences, University of Southern California, Los Angeles, CA 90089, U.S.A.\goodbreak
\and
Department of Physics and Astronomy, University College London, London WC1E 6BT, U.K.\goodbreak
\and
Department of Physics, Florida State University, Keen Physics Building, 77 Chieftan Way, Tallahassee, Florida, U.S.A.\goodbreak
\and
Department of Physics, Gustaf H\"{a}llstr\"{o}min katu 2a, University of Helsinki, Helsinki, Finland\goodbreak
\and
Department of Physics, Princeton University, Princeton, New Jersey, U.S.A.\goodbreak
\and
Department of Physics, University of Alberta, 11322-89 Avenue, Edmonton, Alberta, T6G 2G7, Canada\goodbreak
\and
Department of Physics, University of California, Santa Barbara, California, U.S.A.\goodbreak
\and
Department of Physics, University of Illinois at Urbana-Champaign, 1110 West Green Street, Urbana, Illinois, U.S.A.\goodbreak
\and
Dipartimento di Fisica e Astronomia G. Galilei, Universit\`{a} degli Studi di Padova, via Marzolo 8, 35131 Padova, Italy\goodbreak
\and
Dipartimento di Fisica e Scienze della Terra, Universit\`{a} di Ferrara, Via Saragat 1, 44122 Ferrara, Italy\goodbreak
\and
Dipartimento di Fisica, Universit\`{a} La Sapienza, P. le A. Moro 2, Roma, Italy\goodbreak
\and
Dipartimento di Fisica, Universit\`{a} degli Studi di Milano, Via Celoria, 16, Milano, Italy\goodbreak
\and
Dipartimento di Fisica, Universit\`{a} degli Studi di Trieste, via A. Valerio 2, Trieste, Italy\goodbreak
\and
Dipartimento di Matematica, Universit\`{a} di Roma Tor Vergata, Via della Ricerca Scientifica, 1, Roma, Italy\goodbreak
\and
Discovery Center, Niels Bohr Institute, Blegdamsvej 17, Copenhagen, Denmark\goodbreak
\and
Discovery Center, Niels Bohr Institute, Copenhagen University, Blegdamsvej 17, Copenhagen, Denmark\goodbreak
\and
European Space Agency, ESAC, Planck Science Office, Camino bajo del Castillo, s/n, Urbanizaci\'{o}n Villafranca del Castillo, Villanueva de la Ca\~{n}ada, Madrid, Spain\goodbreak
\and
European Space Agency, ESTEC, Keplerlaan 1, 2201 AZ Noordwijk, The Netherlands\goodbreak
\and
Gran Sasso Science Institute, INFN, viale F. Crispi 7, 67100 L'Aquila, Italy\goodbreak
\and
HGSFP and University of Heidelberg, Theoretical Physics Department, Philosophenweg 16, 69120, Heidelberg, Germany\goodbreak
\and
Helsinki Institute of Physics, Gustaf H\"{a}llstr\"{o}min katu 2, University of Helsinki, Helsinki, Finland\goodbreak
\and
INAF - Osservatorio Astronomico di Padova, Vicolo dell'Osservatorio 5, Padova, Italy\goodbreak
\and
INAF - Osservatorio Astronomico di Roma, via di Frascati 33, Monte Porzio Catone, Italy\goodbreak
\and
INAF - Osservatorio Astronomico di Trieste, Via G.B. Tiepolo 11, Trieste, Italy\goodbreak
\and
INAF/IASF Bologna, Via Gobetti 101, Bologna, Italy\goodbreak
\and
INAF/IASF Milano, Via E. Bassini 15, Milano, Italy\goodbreak
\and
INFN, Sezione di Bologna, viale Berti Pichat 6/2, 40127 Bologna, Italy\goodbreak
\and
INFN, Sezione di Ferrara, Via Saragat 1, 44122 Ferrara, Italy\goodbreak
\and
INFN, Sezione di Roma 1, Universit\`{a} di Roma Sapienza, Piazzale Aldo Moro 2, 00185, Roma, Italy\goodbreak
\and
INFN, Sezione di Roma 2, Universit\`{a} di Roma Tor Vergata, Via della Ricerca Scientifica, 1, Roma, Italy\goodbreak
\and
INFN/National Institute for Nuclear Physics, Via Valerio 2, I-34127 Trieste, Italy\goodbreak
\and
IPAG: Institut de Plan\'{e}tologie et d'Astrophysique de Grenoble, Universit\'{e} Grenoble Alpes, IPAG, F-38000 Grenoble, France, CNRS, IPAG, F-38000 Grenoble, France\goodbreak
\and
IUCAA, Post Bag 4, Ganeshkhind, Pune University Campus, Pune 411 007, India\goodbreak
\and
Imperial College London, Astrophysics group, Blackett Laboratory, Prince Consort Road, London, SW7 2AZ, U.K.\goodbreak
\and
Infrared Processing and Analysis Center, California Institute of Technology, Pasadena, CA 91125, U.S.A.\goodbreak
\and
Institut N\'{e}el, CNRS, Universit\'{e} Joseph Fourier Grenoble I, 25 rue des Martyrs, Grenoble, France\goodbreak
\and
Institut Universitaire de France, 103, bd Saint-Michel, 75005, Paris, France\goodbreak
\and
Institut d'Astrophysique Spatiale, CNRS, Univ. Paris-Sud, Universit\'{e} Paris-Saclay, B\^{a}t. 121, 91405 Orsay cedex, France\goodbreak
\and
Institut d'Astrophysique de Paris, CNRS (UMR7095), 98 bis Boulevard Arago, F-75014, Paris, France\goodbreak
\and
Institut f\"ur Theoretische Teilchenphysik und Kosmologie, RWTH Aachen University, D-52056 Aachen, Germany\goodbreak
\and
Institute for Space Sciences, Bucharest-Magurale, Romania\goodbreak
\and
Institute of Astronomy, University of Cambridge, Madingley Road, Cambridge CB3 0HA, U.K.\goodbreak
\and
Institute of Theoretical Astrophysics, University of Oslo, Blindern, Oslo, Norway\goodbreak
\and
Instituto de Astrof\'{\i}sica de Canarias, C/V\'{\i}a L\'{a}ctea s/n, La Laguna, Tenerife, Spain\goodbreak
\and
Instituto de F\'{\i}sica de Cantabria (CSIC-Universidad de Cantabria), Avda. de los Castros s/n, Santander, Spain\goodbreak
\and
Istituto Nazionale di Fisica Nucleare, Sezione di Padova, via Marzolo 8, I-35131 Padova, Italy\goodbreak
\and
Jet Propulsion Laboratory, California Institute of Technology, 4800 Oak Grove Drive, Pasadena, California, U.S.A.\goodbreak
\and
Jodrell Bank Centre for Astrophysics, Alan Turing Building, School of Physics and Astronomy, The University of Manchester, Oxford Road, Manchester, M13 9PL, U.K.\goodbreak
\and
Kavli Institute for Cosmological Physics, University of Chicago, Chicago, IL 60637, USA\goodbreak
\and
Kavli Institute for Cosmology Cambridge, Madingley Road, Cambridge, CB3 0HA, U.K.\goodbreak
\and
Kazan Federal University, 18 Kremlyovskaya St., Kazan, 420008, Russia\goodbreak
\and
LAL, Universit\'{e} Paris-Sud, CNRS/IN2P3, Orsay, France\goodbreak
\and
LERMA, CNRS, Observatoire de Paris, 61 Avenue de l'Observatoire, Paris, France\goodbreak
\and
Laboratoire AIM, IRFU/Service d'Astrophysique - CEA/DSM - CNRS - Universit\'{e} Paris Diderot, B\^{a}t. 709, CEA-Saclay, F-91191 Gif-sur-Yvette Cedex, France\goodbreak
\and
Laboratoire Traitement et Communication de l'Information, CNRS (UMR 5141) and T\'{e}l\'{e}com ParisTech, 46 rue Barrault F-75634 Paris Cedex 13, France\goodbreak
\and
Laboratoire de Physique Subatomique et Cosmologie, Universit\'{e} Grenoble-Alpes, CNRS/IN2P3, 53, rue des Martyrs, 38026 Grenoble Cedex, France\goodbreak
\and
Laboratoire de Physique Th\'{e}orique, Universit\'{e} Paris-Sud 11 \& CNRS, B\^{a}timent 210, 91405 Orsay, France\goodbreak
\and
Lawrence Berkeley National Laboratory, Berkeley, California, U.S.A.\goodbreak
\and
Lebedev Physical Institute of the Russian Academy of Sciences, Astro Space Centre, 84/32 Profsoyuznaya st., Moscow, GSP-7, 117997, Russia\goodbreak
\and
Max-Planck-Institut f\"{u}r Astrophysik, Karl-Schwarzschild-Str. 1, 85741 Garching, Germany\goodbreak
\and
McGill Physics, Ernest Rutherford Physics Building, McGill University, 3600 rue University, Montr\'{e}al, QC, H3A 2T8, Canada\goodbreak
\and
Mullard Space Science Laboratory, University College London, Surrey RH5 6NT, U.K.\goodbreak
\and
National University of Ireland, Department of Experimental Physics, Maynooth, Co. Kildare, Ireland\goodbreak
\and
Nicolaus Copernicus Astronomical Center, Bartycka 18, 00-716 Warsaw, Poland\goodbreak
\and
Niels Bohr Institute, Blegdamsvej 17, Copenhagen, Denmark\goodbreak
\and
Niels Bohr Institute, Copenhagen University, Blegdamsvej 17, Copenhagen, Denmark\goodbreak
\and
Nordita (Nordic Institute for Theoretical Physics), Roslagstullsbacken 23, SE-106 91 Stockholm, Sweden\goodbreak
\and
Optical Science Laboratory, University College London, Gower Street, London, U.K.\goodbreak
\and
SISSA, Astrophysics Sector, via Bonomea 265, 34136, Trieste, Italy\goodbreak
\and
SMARTEST Research Centre, Universit\`{a} degli Studi e-Campus, Via Isimbardi 10, Novedrate (CO), 22060, Italy\goodbreak
\and
School of Physics and Astronomy, Cardiff University, Queens Buildings, The Parade, Cardiff, CF24 3AA, U.K.\goodbreak
\and
School of Physics and Astronomy, University of Nottingham, Nottingham NG7 2RD, U.K.\goodbreak
\and
Sorbonne Universit\'{e}-UPMC, UMR7095, Institut d'Astrophysique de Paris, 98 bis Boulevard Arago, F-75014, Paris, France\goodbreak
\and
Space Sciences Laboratory, University of California, Berkeley, California, U.S.A.\goodbreak
\and
Special Astrophysical Observatory, Russian Academy of Sciences, Nizhnij Arkhyz, Zelenchukskiy region, Karachai-Cherkessian Republic, 369167, Russia\goodbreak
\and
Stanford University, Dept of Physics, Varian Physics Bldg, 382 Via Pueblo Mall, Stanford, California, U.S.A.\goodbreak
\and
Sub-Department of Astrophysics, University of Oxford, Keble Road, Oxford OX1 3RH, U.K.\goodbreak
\and
The Oskar Klein Centre for Cosmoparticle Physics, Department of Physics,Stockholm University, AlbaNova, SE-106 91 Stockholm, Sweden\goodbreak
\and
Theory Division, PH-TH, CERN, CH-1211, Geneva 23, Switzerland\goodbreak
\and
UPMC Univ Paris 06, UMR7095, 98 bis Boulevard Arago, F-75014, Paris, France\goodbreak
\and
Universit\'{e} de Toulouse, UPS-OMP, IRAP, F-31028 Toulouse cedex 4, France\goodbreak
% ...
}